\documentclass[aps,prl,twocolumn,preprintnumbers,amsmath,amssymb,superscriptaddress]{revtex4-2}
\usepackage[english]{babel}
\usepackage{blindtext}
\usepackage[utf8]{inputenc}
\usepackage{graphicx}
\usepackage{bm}
\usepackage{latexsym}
\usepackage{xcolor}
\usepackage{physics}
\usepackage{amsfonts}
\usepackage{bbold}
\usepackage{scalerel}
\usepackage{appendix}
\usepackage{makecell}
\usepackage{soul}
\usepackage[bookmarks=true,         
            unicode=false,          
            pdftoolbar=true,        
            pdfmenubar=true,        
            pdffitwindow=false,     
            pdfstartview={FitH},    
            pdfsubject={},   
            pdfcreator={},   
            pdfproducer={}, 
            pdfkeywords={} {} {}, 
            pdfnewwindow=true,      
            colorlinks=true,       
            colorlinks = true,
            linkcolor = blue,
            urlcolor  = blue,
            citecolor = blue,
            anchorcolor = blue]{hyperref}

\graphicspath{{figures_PRL/}}
\newcommand{\bk}{\mathbf{k}}

\definecolor{green(html/cssgreen)}{rgb}{0.0, 0.5, 0.0}
\newcommand{\PMB}[1]{#1}

\usepackage{xr}
\makeatletter
\newcommand*{\addFileDependency}[1]{
 \typeout{(#1)}
 \@addtofilelist{#1}
  \IfFileExists{#1}{}{\typeout{No file #1.}}
}
\makeatother

\newcommand*{\myexternaldocument}[1]{
    \externaldocument{#1}
    \addFileDependency{#1.tex}
    \addFileDependency{#1.aux}
}
\myexternaldocument{SM_PRBLetter}
\usepackage{pdfpages}

\makeatletter
\AtBeginDocument{\let\LS@rot\@undefined}
\makeatother
\usepackage{pgf,tikz}
\begin{document}

\title{Interaction-driven first-order and higher-order topological superconductivity}%
\date{\today}

\author{Pietro M. Bonetti}
\thanks{These two authors contributed equally}
\affiliation{Max Planck Institute for Solid State Research, Heisenbergstrasse 1, D-70569 Stuttgart, Germany}
\author{Debmalya Chakraborty}
\thanks{These two authors contributed equally}
\affiliation{Department of Physics and Astronomy, Uppsala University, Box 516, S-751 20 Uppsala, Sweden}
\author{Xianxin Wu}
\affiliation{Institute for Theoretical Physics, Chinese Academy of Sciences, Beijing, China}
\author{Andreas P. Schnyder}
\affiliation{Max Planck Institute for Solid State Research, Heisenbergstrasse 1, D-70569 Stuttgart, Germany}
\affiliation{Yukawa Institute for Theoretical Physics (YITP), Kyoto University, Kyoto 606-8502, Japan}

\begin{abstract}
We investigate topological superconductivity in the Rashba-Hubbard model,
describing heavy-atom superlattice and van der Waals materials with broken inversion.
We focus in particular on fillings close to the van Hove singularities, where
a large density of states enhances the superconducting transition temperature.
To determine the topology of the superconducting gaps and to analyze the stability of their surface states
in the presence of disorder and residual interactions, we 
\PMB{employ} an fRG+MFT approach, which combines the unbiased functional renormalization group (fRG) with a real-space mean-field theory (MFT).
Our approach uncovers a cascade of topological superconducting states,
including $A_1$ and $B_1$ pairings, whose wave functions are of dominant $p$- and $d$-wave character, respectively, as well as a time-reversal breaking  $A_1 + i B_1$ pairing.
While the $A_1$ and $B_1$ states have first order topology with helical and flat-band Majorana edge states, respectively, the $A_1 + i B_1$ pairing exhibits
second-order topology with Majorana corner modes. We investigate the disorder stability
of the bulk superconducting states,
analyze interaction-induced instabilites of the edge states, and discuss 
implications for experimental systems.
%
\end{abstract}
\pacs{}
\maketitle
%

Topological superconductors (TSCs) are of high current interest due to their exceptional properties and
potential for applications in quantum information technologies~\cite{beenakker_review_13,Schnyder_review_2015,chiu_RMP_16,Sato_review_2017,Sharma_review_2022,vonrohr2023chemical,flensberg_stern_review_21}. A variety of heterostructures~\cite{flensberg_stern_review_21} and candidate materials~\cite{jiangping_hu_review_18,WuXX2021,iron_based_TSC_nat_phys_15,doi:10.1126/science.aao1797}, where topological superconductivity is expected to occur, have been investigated. However, despite tremendous efforts, the ideal topological superconductor suitable for the envisioned applications has yet to be found. Two major obstacles in this research field are small superconducting gaps and omnipresent disorder. Since the topology only protects against perturbations smaller than the superconducting gap, it is of paramount importance to find intrinsic topological superconductors with larger gaps compared to proximty-induced superconductors~\cite{Mourik12,Aghaee23}.

Finding intrinsic topological superconductivity needs superconductivity to be originating from electron-electron repulsive interactions since electron-phonon interactions commonly only give conventional non-topological superconductivity~\cite{marsiglio2001electron,PhysRevB.90.184512}. 
One possible path to high-$T_c$ topological superconductivity is to search for platforms using high-$T_c$ cuprate superconductors, for example as proposed in twisted bilayer cuprates \cite{Can21}. However, the lack of tunability in material control of cuprates has made it hard to experimentally realize twisted bilayers of cuprates. An alternative strategy investigated in recent times is to use a large density of states at the Fermi level \cite{Kohn65}. This strategy has been envisaged in material candidates where the Fermi level lies close to a van Hove singularity \cite{Nandkishore12} or near flat bands \cite{Gonzalez19}. But a major hindrance towards such possibility of superconductivity is that a large density of states is also associated with particle-hole orders, like ferromagnetism, due to the Stoner criterion \cite{blundell2001magnetism} or spin/charge density waves due to nesting \cite{Aaron19,Fleischmann20}. 
However, doping away from a ferromagnetic phase to its critical point can result in triplet topological superconductivity, due to ferromagnetic fluctuations \cite{Fay80}.
But the resulting triplet superconductivity is highly fine-tuned,
exists only in a narrow region of
parameter space, and has an exponentially 
small $T_c$, due to the large distance from the singular density of states \cite{Honerkamp2001}.

Another ingredient which is often believed to be crucial for topological superconductivity is spin-orbit coupling (SOC). SOC is known to split one van Hove singularity into two and also changes the topological nature of the Fermi surfaces. Hence, the presence of spin-orbit coupling gives the possibility of having high density of states in a large parameter space and additionally gives non-trivial topological band structures, providing a possible route to intrinsic topological superconductivity with large $T_c$. With recent progress in experimental techniques, it has now become possible to fabricate 2D van der Waals materials~\cite{review_2D_mat_novoselov_16,review_nat_van_der_waals_16,review_van_der_Waals_2021} with high tunability both in doping or filling and spin-orbit coupling \cite{shimozawa_superlattice_PRL_14,Liu20,Zhang20,Kang21}.

Motivated by this, we theoretically analyze the  
conditions under which topological superconductivity 
emerges in the Rashba-Hubbard model~\cite{Greco2018,schnyder_greco_ferro_fluctuations_PRB_2020,PhysRevB.99.115122,Wolf2020,yanase_Rashba_Hubbard_PRB_2020,biderang_sirker_PRB_22,kawano_hotta_PRB_23,rashba_hubbard_rachel_PRB_23} in 2D, to capture the simultaneous role of electronic correlations driven by Hubbard interaction, SOC within Rashba model, and singular density of states at van Hove singularities in 2D. Theoretically investigating interactions and singular density of states is challenging, since methods like mean-field or random phase approximation calculations cannot capture the mutual interference between particle-particle and particle-hole instabilities at singularities~\cite{Schulz87}, 
and Quantum Monte Carlo applied to the
Rashba-Hubbard model is plagued by the sign problem. We therefore employ the functional renormalization group (fRG)~\cite{Kopietz2010,Metzner2012,Lichtenstein2017},
which treats all instability channels
on an equal footing, and augment it with a mean-field theory (MFT) in order to capture the nature of the superconductivity deep inside the phase.

Specifically, we find that while magnetism is suppressed by SOC, magnetic fluctuations near the van Hove singularities lead to $A_1$ and $B_1$ superconducting states, with dominant $p$-  and $d$-wave like pairing character, respectively, in a large parameter regime. Both of these pairing states exhibit nontrivial first order topology~\cite{chiu_RMP_16} and host helical and flat-band Majorana edge states, respectively. 
Remarkably, we also find near the phase boundary between these two states a time-reversal breaking $A_1 + i B_1$ pairing state with higher-order topology and Majorana corner modes~\cite{wang_d_plus_ip_PRB_18,yan_wang_high_order_SC_PRL_18,wang_fan_PRL_18,WuXXNSR,hsu_sau_PRL_20,marsiglio_higher_order_SC_PRL_20,PhysRevX.10.041014,ikegaya_corner_modes_PRR_21,Li_2D_materials_2022}. Furthermore, we extend the fRG+MFT method to real-space, which allows us to deduce the topological and edge properties of the superconducting states, as well as the (in-)stabilities of the edge states against residual interactions, and the stability of these states to disorder. Our fRG+MFT approach in real space reveals that while both the helical Majorana and corner edge states are robust to residual interactions, the flat-band Majorana states are unstable towards the formation of a 1D phase crystal~\cite{Hakansson2015,Holmvall20,Chakraborty22}. We further find that the Majorana corner modes in the $A_1 + i B_1$ pairing state remain robust to disorder, defying the usual expectation that the sign-changing $B_1$ pairing, responsible for the corner modes, is sensitive to disorder.

\noindent
\textit{Model and methods.--}
We start from the
Rashba-Hubbard Hamiltonian on the square lattice 
given by
$H=H_{0}+H_{U}$,  with
\begin{equation}\label{eq: Rashba Hubbard hamiltonian}
    \begin{split}
        &H_{0} = \sum_{j,j',\sigma} t_{jj'} c^\dagger_{j,\sigma} c_{j',\sigma}-\mu\sum_{j,\sigma}n_{j,\sigma}\\
        &+ i\sum_{\substack{j,j'\\\sigma,\sigma'}} \lambda_{jj'} \left[\left(\mathbf{r}_j-\mathbf{r}_{j'}\right)\times c^\dagger_{j,\sigma}\vec{\tau}_{\sigma\sigma'}c_{j',\sigma'}\right]_{z}+\textrm{H.c.},\\
        &H_{U}=U\sum_{j}n_{j,\uparrow}n_{j,\downarrow},
    \end{split}
\end{equation}
where $c_{j,\sigma}$ ($c_{j,\sigma}^\dagger$) is the annihilation (creation) operator of an electron at site $j$ with spin projection $\sigma$, 
$\vec{\tau}$ are the Pauli matrices, $\mathbf{r}_j$ is the lattice coordinate of site $j$,
$n_{j,\sigma}=c^\dagger_{j,\sigma}c_{j,\sigma}$ is the spin-resolved particle number operator, $\mu$ is the chemical potential. 
%
%
In the following, we choose the hopping amplitudes such that $t_{\langle j,j'\rangle}=-t$ when $j$ and $j'$ are nearest neighbors, $t_{\langle\langle j,j'\rangle\rangle}=-t'$ when $j$ and $j'$ are second neighbors, and zero otherwise. We set $t$ as the energy unit. We also consider the Rashba 
SOC to be nonzero only for nearest neighbors $\lambda_{\langle j,j'\rangle}=\lambda t$. In the rest of the paper, we take $\lambda=0.3$. Results for different values of $\lambda$ are shown 
in the Supplemental Material (SM)~\cite{sm_note}.

\ifx
Fourier transforming the quadratic part of the Hamiltonian, one obtains
\begin{equation}
    H_0 = \int_\bk c^\dagger_\bk \hat{\mathcal{H}}_\bk c_\bk,
\end{equation}
where $\int_\bk=\int_\mathrm{BZ} \frac{d^2k}{(2\pi)^2}$ is a shorthand for the integral over the first Brillouin zone. $\hat{\mathcal{H}}_\bk$ is a $2\times2$ matrix that reads as
\begin{equation}\label{eq: H0k}
    \hat{\mathcal{H}}_\bk = (\epsilon_\bk-\mu)\mathbb{1}+\lambda\vec{g}_\bk \cdot\vec{\tau},
\end{equation}
with $\epsilon_\bk=-2t(\cos k_x+\cos k_y)-4t'\cos k_x\cos k_y$ the Fourier transform of $t_{jj'}$, and $\vec{g}_\bk=2t(-\sin k_y,\sin k_x,0)$. Hamiltonian \eqref{eq: H0k} can be easily diagonalized introducing the so-called \emph{helicity basis}, $c_\bk \to U_\bk c_\bk$, with
%
    \begin{align}
        & U_\bk = \frac{1}{\sqrt{2}}\left(
        \begin{array}{cc}
            1 & e^{-i\phi_\bk} \\
            -e^{i\phi_\bk} & 1
        \end{array}\right),
    \end{align}
%
with $e^{i\phi_\bk}=(g_{1,\bk}+ig_{2,\bk})/|\vec{g}_\bk|$, leading to the quasiparticle bands
%
 $ E^\pm_\bk = \epsilon_\bk - \mu \pm \lambda |\vec{g}_\bk| $. 
%
\fi  

%
\begin{figure}
    \centering
    \includegraphics[width=0.5\textwidth]{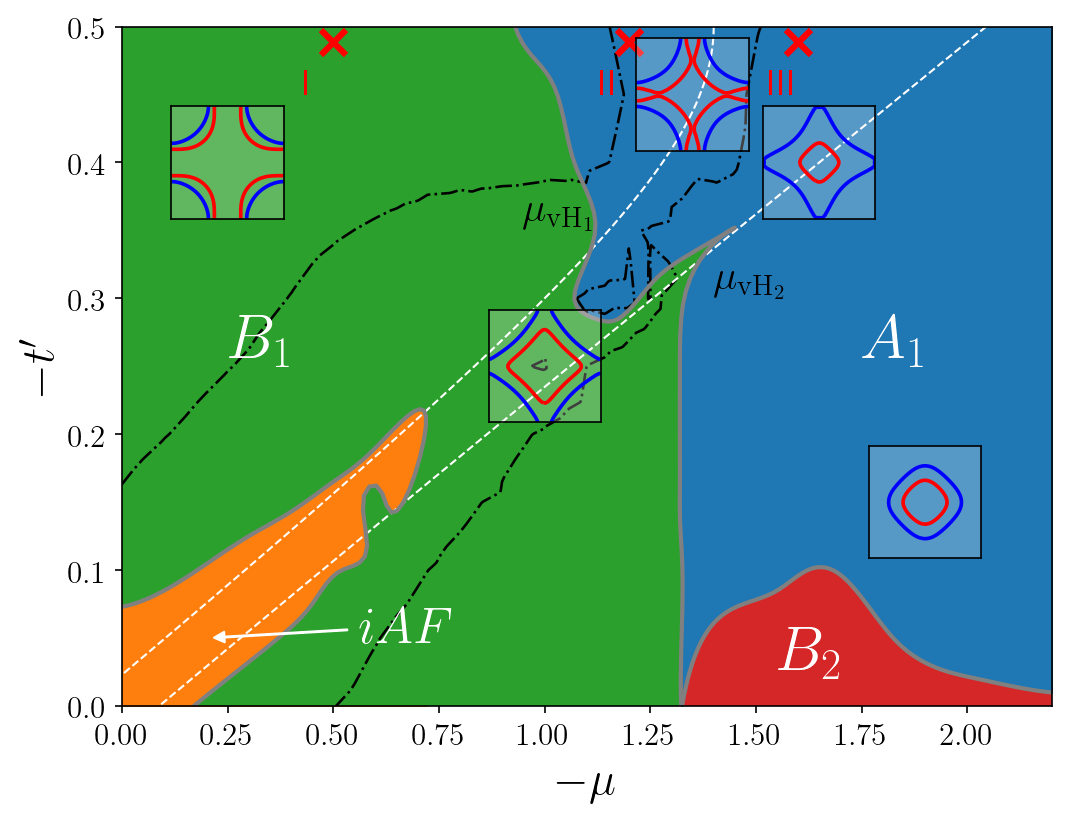}
    \caption{
Phase diagram of the Rashba-Hubbard model on the square lattice as a function of chemical potential $\mu$ and second-neighbor hopping~$t'$ for $U=3t$. The dashed lines indicate the values of the chemical potential $\mu_\mathrm{vH_1}(t')$, $\mu_\mathrm{vH_2}(t')$ where the van Hove singularities VHS1 and VHS2 occur. 
The dashed-dotted lines
enclose the region where $T_c$ exceeds $10^{-6}t$.}
    \label{fig: phase diag}
\end{figure}

We investigate the superconducting (SC) phases of the model in Eq.~\eqref{eq: Rashba Hubbard hamiltonian} by means of the functional renormalization (fRG) group, combined with mean-field theory (fRG+MFT)~\cite{Wang2014,Yamase2016,Vilardi2020,Bonetti2020}. This results in a renormalized superconducting gap equation of the form
\begin{equation} \label{eq: gap equation main}
    \Delta^{\sigma\sigma'}_\bk = \sum_{s,s'}\int_{\bk'} \widetilde{V}^{\sigma\sigma'ss'}_{\bk\bk'} \left[T\sum_n F^{ss'}(\bk',\nu_n)\right],
\end{equation}
where $\nu_n$ are fermionic Matsubara frequencies,
$T$ is the temperature, 
$\Delta^{\sigma\sigma'}_\bk$  is the SC order parameter, and $F^{ss'}(\bk',\nu_n)$ is the spin-resolved anomalous propagator, describing the propagation of a hole that gets reflected into a particle or vice-versa. The function $\widetilde{V}^{\sigma\sigma'ss'}_{\bk\bk'}$ describes an effective interaction, computed by means of the fRG 
(see SM for details~\cite{sm_note}).  
%
%
To determine the symmetry (and not the size) of the SC gap, we linearize Eq.~\eqref{eq: gap equation main} in $\Delta^{\sigma\sigma'}_\bk$. The equation becomes therefore an eigenvalue problem, where the eigenvector corresponding to the largest positive eigenvalue gives the information on the symmetry of the 
leading superconducting state. 
%

\noindent
\textit{Phase diagram.--}
In Fig.~\ref{fig: phase diag}, we show the superconducting phase diagram as a function of  chemical potential $\mu\leq0$ and second-neighbor hopping $t'\leq0$. We also show the different topologies of the two Fermi surfaces,
as well as the lines along which the van Hove singularities of the two quasiparticle bands reach the Fermi level. 
We note that the density of states and position in $k$ space of the two van Hove singularities are marked differently: VHS1 (left line) has a larger density of states than VHS2 (right line) and is also located further away from the $(\pi, 0)$ point, 
see SM~\cite{sm_note}. In the orange region of Fig.~\ref{fig: phase diag}, labeled as iAF, the fRG flow signals an (incommensurate) antiferromagnetic instability. In the region enclosed by the dashed-dotted black lines the superconducting transition temperature exceeds $T=10^{-6}t$, which is the lowest temperature accessible in our fRG computation.

A large portion of the phase diagram displays a leading superconducting state living in the $B_1$ representation of the combined spin-lattice symmetry group of Hamiltonian~\eqref{eq: Rashba Hubbard hamiltonian} 
~\cite{sm_note}. The resulting gap function can to leading order be expressed as 
\begin{equation*}\label{eq: B1 gap}
        \alpha_s (\cos k_x - \cos k_y ) t^0 + \alpha_t \left(\sin k_y t^1 + \sin k_x t^2\right),
\end{equation*} 
where $t^\mu=i\tau^\mu\tau^2$, and $\alpha_s$ and $\alpha_t$ are free parameters. Note that the singlet component of the gap (the one proportional to $t^0$) is the $d_{x^2-y^2}$-wave SC order parameter one would obtain in the Hubbard model without introducing  SOC. Higher-order harmonics are considered in our calculations, but the effective attractions in the higher-harmonic channels are found to be very small.
At larger absolute values of the chemical potential, we obtain a phase belonging to the $A_1$ representation of the discrete symmetry group of the Hamiltonian. The SC gap in this state is a superposition of an extended $s$-wave in the singlet component and a helical $p$-wave in the triplet:
\begin{equation*} \label{eq: A1 gap}
    \alpha_s (\cos k_x + \cos k_y ) t^0 + \alpha_t \left(-\sin k_y t^1 + \sin k_x t^2\right). 
\end{equation*}

For larger $-\mu$ and small $-t'$, we also find a $B_2$ phase, characterized, however, by rather small values of the leading eigenvalue, resulting in low transition temperatures. The gap function in this phase is given by \PMB{a dominant singlet component in the $d_{xy}$ symmetry channel, and a subdominant triplet part}.
%
%
We note that near the
boundary between differently colored regions in Fig.~\ref{fig: phase diag}
there is the possibility of a coexistence phase that combines the two order parameters (see discussion below). Hence, in general there will be two 
transitions
as one goes from, e.g., deep inside the $B_1$ state to the $A_1$ state.
Finally, we note that SOC disfavors ferromagnetic phases expected near van Hove singularities~\cite{Honerkamp2001}. Instead, spin fluctuations drive the formation of the SC phases in a large parameter regime even at van Hove fillings, see SM~\cite{sm_note}.

\begin{figure}
    \centering
    \includegraphics[width=.5\textwidth]{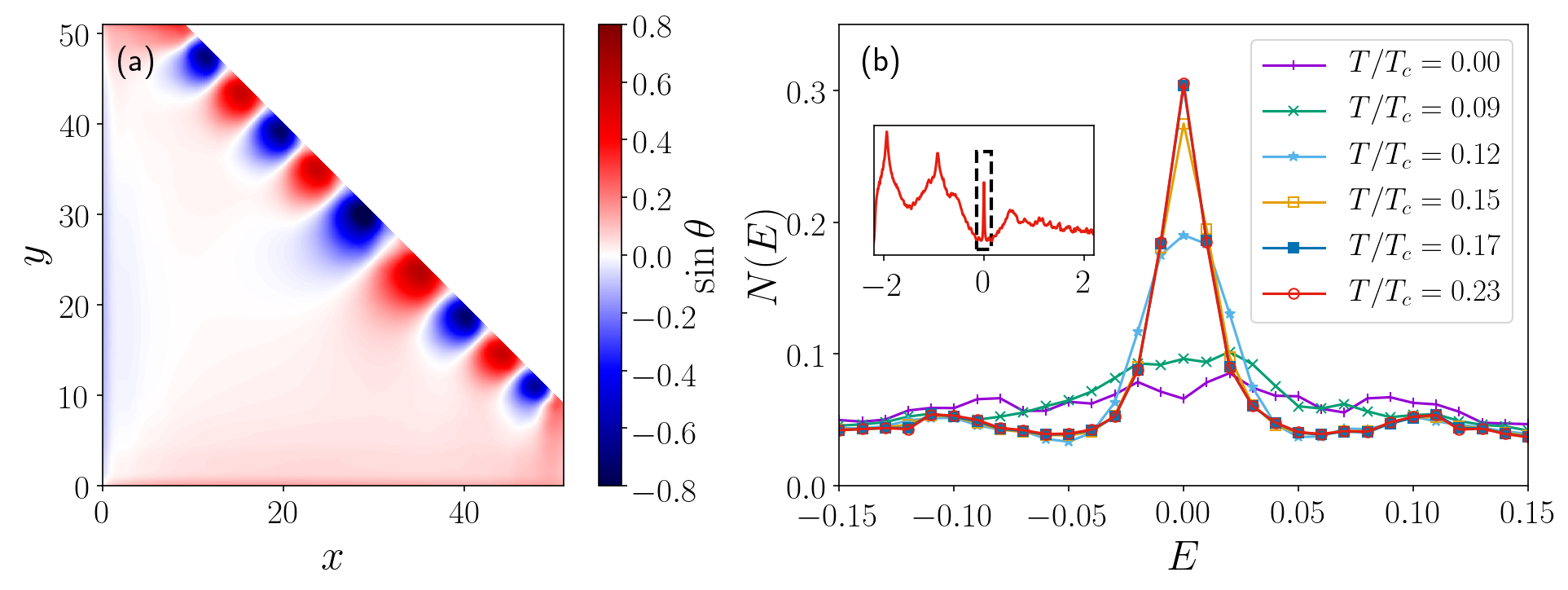}
\caption{(a) Phase crystal forming at the (11) edge. The sine of the phase $\theta$ of the $d$-wave superconducting order parameter is plotted in real space at $T=0$
with the bulk value of $\sin \theta$ being subtracted. At the (11) edge there are clear modulations visible in $\sin \theta$.
(b) Density of states $N(E)$ around zero energy for different temperatures $T/T_c$. The inset shows the
density of states at $T/T_c=0.23$ in a wider range.}
    \label{fig: fig3}
\end{figure}

Having identified different superconducting phases in Fig.~\ref{fig: phase diag}, we then focus on each of them individually by investigating 
the characteristic points marked by red crosses in Fig.~\ref{fig: phase diag}. Additionally, we extend the fRG+MFT method to real-space to investigate the disorder
stability of the phases and to determine possible interaction-induced edge instabilities~\cite{sm_note}. Furthermore, we scale the interaction strengths by 10 in order to \PMB{render the investigation} of the edge properties \PMB{on finite size systems computationally amenable}.

\noindent
\textit{B$_1$ pairing state.--}
In Fig.~\ref{fig: fig3}, we investigate the properties of the $B_1$ phase for the parameters marked by 
cross~I ($\mu=-0.5$ and $t'=-0.5$) in Fig.~\ref{fig: phase diag}. 
In this regime, the gap functions show (quasi-) nodes along the diagonal $k_x=\pm k_y$, see SM~\cite{sm_note}. This nodal structure suggests that a (11) edge is pair-breaking \cite{Kashiwaya00,Lofwander01} and will form a flat band of topological zero-energy states \cite{Ryu02,Sato11,Schnyder15,schnyder_brydon_PRL_13} protected by time-reversal symmetry and translational symmetry along (11). 
The large degeneracy of these zero-energy states makes them thermodynamically unstable and prone to symmetry breaking. To investigate the edge properties, we consider an open boundary geometry (see Fig.~\ref{fig: fig3}(a)) with one (11) edge. As shown in (a), at low temperatures, the (11) edge hosts oscillations in the phase $\theta$ of the $d$-wave superconducting order parameter, called phase crystals breaking both time-reversal symmetry and translational symmetry along the (11) edge \cite{Hakansson2015,Holmvall20,Chakraborty22}. 
In comparison to earlier literature on phase crystals, one remarkable feature of the phase crystals obtained here is that they are robust to the additional presence of a subdominant triplet superconducting order parameter originating in the bulk 
due to SOC. We calculate the spatially averaged density of states $N(E)=1/N\sum_{i,n,\sigma}|u_{i\sigma}^{n}|^2\delta(E-E_n)+|v_{i\sigma}^{n}|^2\delta(E+E_n)$, where $N$ is the total number of lattice sites, and $u_{i\sigma}^{n}$ and $v_{i\sigma}^{n}$ are the eigenfunctions with eigenvalues $E$. To numerically evaluate $N(E)$, we use a Lorentzian with fixed width 0.01 to calculate the delta-function. 
As shown in Fig.~\ref{fig: fig3}(b), $N(E)$ shows a large zero-bias peak for $T/T_c>0.17$, showing the presence of a flat band of zero-energy states, which does not change with increasing temperature. 
Due to the formation of the phase crystal at $T/T_c\approx0.17$, the zero-bias peak gets suppressed for lower temperatures since the phase crystals Doppler shifts the zero-energy states to finite energies. With lowering temperature, the shift increases. 


\noindent
\textit{A$_1$ pairing state.--}
We now investigate the topological edge states of the $A_1$ phase for the parameters marked by cross III ($\mu=-1.6$ and $t'=-0.5$) in Fig.~\ref{fig: phase diag}. This $A_1$ pairing is a time-reversal symmetric fully gapped superconductor and belongs to the DIII class, characterized by a $\mathbb{Z}_2$ invariant in two dimension~\cite{chiu_RMP_16}. 
In our case, there is one pocket around the M point with negative pairing and the system is topologically nontrivial according to $N_{2D}=\Pi_{s}[\text{sgn}(\Delta_s)]^{m_s}$, where $m_s$ is the number of time-reversal invariant points enclosed by the $s$th Fermi surface~\cite{Qi2010}. Due to this non-trivial topology, we find helical edge states with open boundaries, shown in Sec.~SVII of the SM~\cite{sm_note}.

\begin{figure}
    \centering
    \includegraphics[width=.5\textwidth]{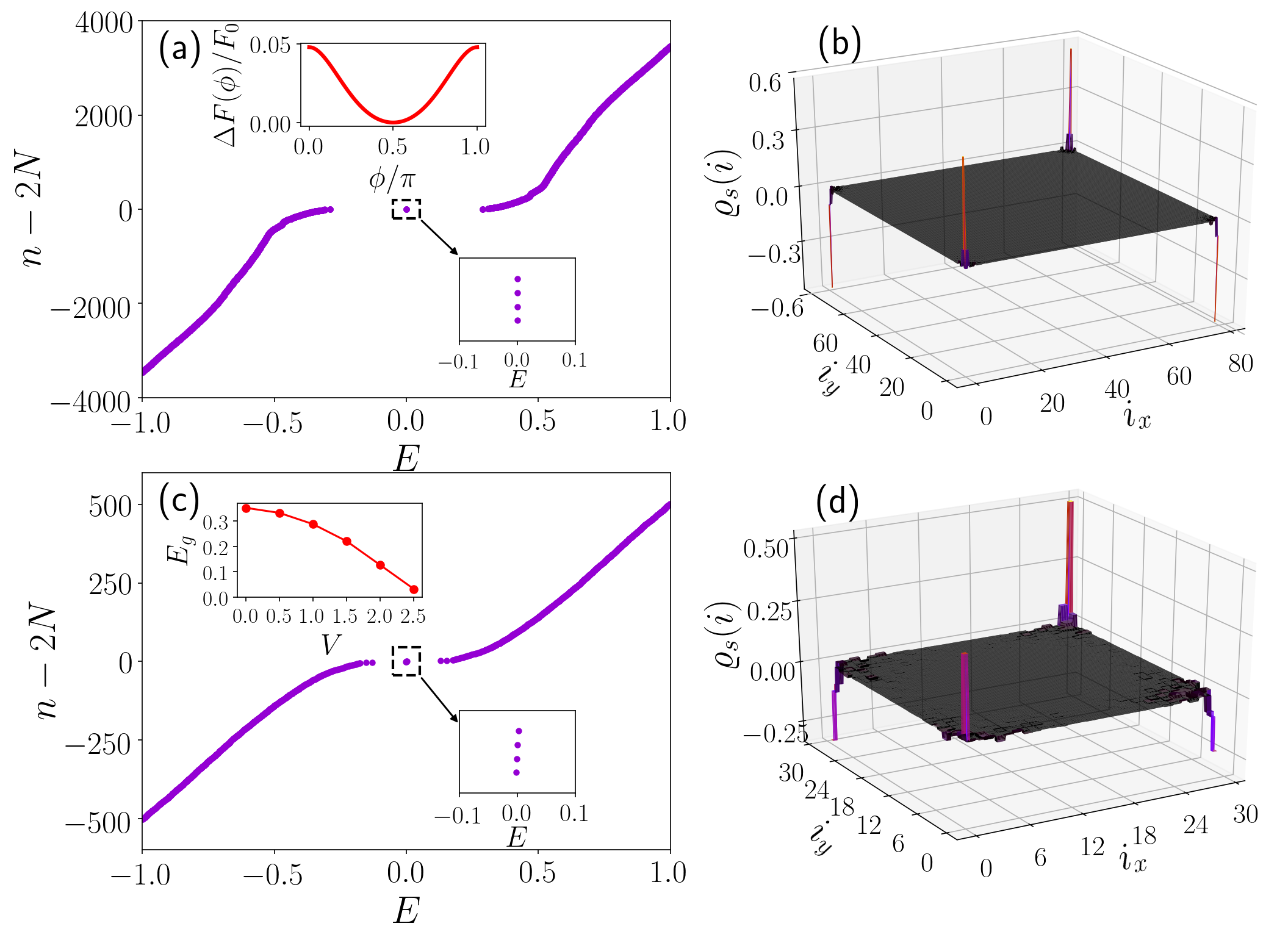}
\caption{(a) Eigenvalues with open boundary conditions along both $x$ and $y$, showing four Majorana states. Upper inset: Normalized condenstation energy $\Delta F$ as a function of the relative phase $\phi$ between $A_{1}$ and $B_{1}$ order parameters. Lower inset: zoomed view of the four Majorana states. 
(b)~Spin projected wave function of the four Majorana states. 
(c,d) Same as (a,b) but with disorder $V=1.5$. Upper inset of (c): Bulk energy gap $E_g$ as a function of disorder strength $V$. Here, $T=0$.}
    \label{fig: fig5}
\end{figure}

\noindent
\textit{A$_1+ i $B$_1$ pairing state.--}
The transition from the $B_1$ superconducting state to the $A_1$ state with increasing $|\mu|$ in Fig.~\ref{fig: phase diag} gives the possibility of a coexisting phase where both order parameters are comparable. To explore this possibility, we fix the parameters to the values marked by cross II ($\mu=-1.2$ and $t'=-0.5$) in Fig.~\ref{fig: phase diag}, which is near the phase boundary.
Interestingly, we find a time-reversal symmetry breaking $A_1+iB_1$ superconducting phase as the lowest energy state. 
To verify that the relative phase of $\pi/2$ between the $A_1$ and the $B_1$ order parameters is indeed the global minimum, we compare the condensation energies of the superconducting states with different relative phase $\phi$. As shown in the inset of Fig.~\ref{fig: fig5}(a), $\phi=\pm\pi/2$ gives the largest condensation energy and consequently $A_1\pm iB_1$ is lowest in energy. 
Due to the imaginary $B_1$ pairing component, time-reversal symmetry and four-fold rotational symmetry $C_4$ are broken in this pairing state, but their combination is preserved, leading to a gap opening in the helical Majorana edge states. 
However, owing to the sign change of the $B_1$ pairing under $C_4$, the mass terms for adjacent edges [(10) and (01) edges] have opposite sign. Therefore, when these two edges meet at the corners, the mass term vanishes and Majorana corner modes are generated, realizing a second-order topological superconductor~\cite{wang_d_plus_ip_PRB_18}, see SM~\cite{sm_note}. 
To demonstrate this nontrivial topology, we study the corner states in a geometry with open boundary conditions along both $x$- and $y$-directions. 
Similar to the $A_1$ phase, we only consider non self-consistent eigenstates with open boundaries taking the fully self-consistent bulk solutions. Remarkably, we find four Majorana zero-energy states located in the superconducting gap, as shown in Fig.~\ref{fig: fig5}(a). 
The wavefunctions corresponding to these Majorana states are localized at the four corners (not shown). 
Spin characteristics of these Majorana states can be visualized by looking at the spin projected wavefunctions $\varrho_{s}(i)=\sum_{\sigma n'}\mathrm{sgn}(\sigma)\left(|u_{i\sigma}^{n'}|^2+|v_{i\sigma}^{n'}|^2\right)$, where $n'$ are the four zero-energy states. In Fig.~\ref{fig: fig5}(b), we see that $\varrho_{s}(i)$ is localized at the four corners with alternating signs for alternating corners. We have also verified the presence of Majorana corner states in a self-consistent calculation with edges for smaller system sizes. Therefore, the Majorana corner states in our numerical calculations confirm that the $A_1+iB_1$ state is a higher-order topological superconductor. 

To analyze the stability of the $A_1+iB_1$ state to perturbations, we study the disorder effects on this phase. We introduce non-magnetic chemical potential disorder by adding a term $H_{V}=\sum_{i}V_i n_{i}$ to the Hamiltonian, with $V_{i}$ being a non-magnetic impurity potential drawn from a random distribution, such that $V_{i}\in [-V/2,V/2]$ uniformly
(i.e., Anderson disorder), and perform a fully self-consistent calculation. As shown in the inset of Fig.~\ref{fig: fig5}(c), the bulk gap $E_g$ reduces with increasing disorder strength $V$, but remains finite for realistic disorder strengths of $V\le 2.0$. Notably, the average order parameters, shown in Fig.~S4(c) of the SM~\cite{sm_note}, show more robust behavior. The disorder-robust behavior of $A_1+iB_1$ is remarkable since the broken time-reversal symmetry makes Anderson's theorem \cite{Anderson59} not applicable, and might be related to the presence of interactions \cite{Garg08,Chakraborty14,Chakraborty17a,Ghosal18}. We also look at the corner states. Fig.~\ref{fig: fig5}(c) and (d) show the eigenvalues and the spin projected wavefunction of the lowest energy states for $V=1.5$. The Majorana corner states persist even in the presence of disorder, showing the stability of the higher-order topological superconductor. This finding is also striking from the following point of view. It is known that topological nature of first-order topological phases make Majoranas survive moderately strong disorder~\cite{queiroz_PRB_14,queiroz_PRB_15,PhysRevB.102.224510}. However, the corner modes in the higher-order $A_1+iB_1$ arise due to the sign change of $B_1$ pairing at adjacent edges. Now, it is commonly believed that $B_1$ pairing is sensitive even to non-magnetic disorder \cite{Alloul09}. Hence, the persistence of Majorana corner modes in the presence of disorder is highly non-intuitive and opens a new way of understanding sensitivity of higher-order topological superconductors. It will also be interesting to investigate in the future the effects of other models of disorder.

 
\noindent
\textit{Discussion.--}
We have shown that the combined effects of van Hove singularities, Rashba SOC, and repulsive Hubbard interactions give rise to a cascade of topological superconducting states, including a nodal $B_1$ state ($d$-wave like) with flat-band Majorana edge modes, a fully gapped $A_1$ state ($p$-wave like) with helical Majorana edge modes, and a time-reversal breaking $A_1 + i B_1$ state with Majorana corner modes, which is also disorder-robust. 
%

\ifx
While the first two pairing states have first order topology, characterized by a winding number and a Pfaffian ${Z}_2$ invariant, respectively, the third one exhibits higher-order topology.
Since Rashba SOC breaks inversion symmetry, all three phases consist of a mixture of spin-singlet and spin-triplet Cooper pairs, whose relative ratio, however,
is vastly different: The $B_1$ and $A_1$ states have dominant singlet and triplet pairing, respectively, whereas the  $A_1 + i B_1$ is composed of an almost equal amount
of singlet and triplet Cooper pairs.
The pairing glue for these superconductors is provided by ferro- and antiferromagnetic spin fluctuations,
which are strongly enhanced near the van Hove fillings. 
Interestingly, the Rashba SOC splits the van Hove singularity at the $X$ point of the square lattice 
into two \emph{different types}, namely, VHS1 which shifts along the $X$-$\Gamma$ direction and
VHS2 which is moved in the $X$-$M$ direction, see SM.
%
%
Remarkably, the divergence at VHS1 is considerably stronger than the one at VHS2, implying that
the higher-order  topological superconductor has the largest $T_c$. 
As a consequence, correlated Rashba materials whose Fermi level can be tuned to VHS1 are good candidates
for intrinsic higher-order topological superconductors. 
\fi

It is remarkable that we find interaction-driven topological superconducting phases in the vicinity of van Hove singularities due to the presence of SOC. We emphasize that the topological superconducting phases emerge robustly in the Rashba-Hubbard model, independent of the size of the Rashba SOC and survive the inclusion of further neighbor hoppings. The size of the Rashba SOC however decides the doping window where topological superconductivity is obtained. Although we have presented here only results for the square lattice, we expect these topological superconductors
to arise also in other 2D lattices, e.g., the triangular or the honeycomb lattice, although with modified irreps,
due to the different spin-lattice symmetry groups. 
While we have only considered effects of a Hubbard onsite interaction, longer ranged interactions can bring in additional interesting aspects, which will be investigated in a future work. For example, on kagome lattice systems, nearest-neighbor interactions play a crucial role due to the sublattice interference effects \cite{Kiesel12,Wu21}. 
%
%

Our results provide a guide to understand and design topological superconductivity in heavy-atom superlattices~\cite{Shimozawa_review_Kondo_superlattice_2016} and van der Waals materials~\cite{review_2D_mat_novoselov_16,review_nat_van_der_waals_16,review_van_der_Waals_2021}. The high variability of these materials may allow to tune the Fermi level to the van Hove fillings, such that an intrinsic topological superconductor with  large $T_c$ can be realized. For example, in LAO/STO~\cite{LAO_STO_triscone_nature} or EuO/KTO~\cite{EuO_KTO_science_21} it is possible to tune the carrier density, and therefore the Fermi surface, by electric gating. With the recent progress in experimental techniques, the strength of the SOC can also be highly tuned. It can either be tuned by applying an electric field \cite{shimozawa_superlattice_PRL_14,Liu20,Zhang20,Kang21} or by changing the geometry of a superlattice. For example,
SOC in the CeCoIn$_5$/YbCoIn$_5$
superlattices can be adjusted by the width of the YbCoIn$_5$ blocks~\cite{matsuda_superlattice_Nature_11,shimozawa_superlattice_PRL_14}. In CeCoIn$_5$/YbCoIn$_5$ superlattices~\cite{matsuda_superlattice_Nature_11,shimozawa_superlattice_PRL_14}, there are already prospective signatures of topological
superconductivity below $T_c \simeq 2$~K, a high value compared to topological superconductivity proposed in semiconductor-superconductor nanowire devices \cite{Mourik12,Aghaee23}. Other promising candidates can be van der Waals heavy-atom materials~\cite{heavy_fermion_van_der_Waal_npj_22,PhysRevLett.84.4986,aps_abstract_heavy_fermion_van_der_Waals}. All these experimental developements and our findings taken together may open up a route to intrinsic topological superconductors being used for the design
of quantum information devices. 
Moreover, the $A_1 + i B_1$ superconductor may show interesting diode~\cite{Ando20,Daido22} and piezoelectric  effects~\cite{PhysRevB.105.024509}, 
since it breaks both inversion and time-reversal symmetry.





\vspace{0.2cm}

\acknowledgements
 
We are thankful to A.~Greco and M.~Hirschmann for enlightening discussions.
D.C. and A.P.S.~thank the Max-Planck-Institute FKF Stuttgart and the YITP Kyoto, respectively, for hospitality during the course of this project. D.C. acknowledges financial support from Kungl. Vetenskapsakademien and C.F. Liljewalchs stipendiestiftelse Foundation.

 

\bibliography{main_PRL.bib}

\newpage


\section*{Supplementary material}

# Supplementary Information - van Hove, Rashba, and Hubbard meet to form first-order and higher-order topological superconductors

This Supplementary Information contains six sections. In SI we discuss the details of the truncated unity renormalization group procedure and explain how it is combined with a momentum-space mean-field theory. In SII we present the symmetry classification of the superconducting states in terms of the irreps of the spin-lattice group. In SIII we provide the details of the real space renormalized mean-field theory and show how it is matched to the fRG. In SIV effective models for the van Hove singularities are derived and analyzed. In SV the corner modes of the $A_1 + iB_1$ pairing state are discussed. The effects of disorder are studied in SVI.

## SI. FRG+MFT METHOD

In this section we provide some details on the fRG+MFT method, employed to determine the leading superconducting symmetry in the Rashba-Hubbard model.

### A. Functional renormalization group

In order to determine the system's instabilities and the leading superconducting symmetry, we run a truncated functional renormalization group flow [32, 33]. The flow is implemented by a progressive integration of fermionic modes by introducing a flowing cutoff in the bare propagator. Here, we choose the so-called *temperature flow* [68]. By rescaling the Grassmann fields according to

$$\eta_{\mathbf{k},\sigma} = T^{\frac{3}{4}} \psi_{\mathbf{k},\sigma}, \tag{S1a}$$

$$\overline{\eta}_{\mathbf{k},\sigma} = T^{\frac{3}{4}} \overline{\psi}_{\mathbf{k},\sigma}, \tag{S1b}$$

where $T$ is the temperature, we obtain the action

$$\mathcal{S}^T[\eta, \overline{\eta}] = \sum_n \int_{\mathbf{k}} \overline{\eta}_{\mathbf{k},n} \left[ T^{-\frac{1}{2}} \left( i\nu_n - \hat{\mathcal{H}}_{\mathbf{k}} \right) \right] \eta_{\mathbf{k},n} + U \sum_{n_1,n_2,n_3} \int_{\mathbf{k}_1,\mathbf{k}_2,\mathbf{k}_3} \overline{\eta}_{\mathbf{k}_1,n_1,\uparrow} \overline{\eta}_{\mathbf{k}_2,n_2,\downarrow} \eta_{\mathbf{k}_4,n_4,\downarrow} \eta_{\mathbf{k}_3,n_3,\uparrow},$$

where $n$ labels fermionic Matsubara frequencies $\nu_n = (2n+1)\pi T$ ($n \in \mathbb{Z}$), $\mathbf{k}_4 = \mathbf{k}_1 + \mathbf{k}_2 - \mathbf{k}_3$ and $n_4 = n_1 + n_2 - n_3$, and $\hat{\mathcal{H}}_{\mathbf{k}}$ is defined in Eq. (3). We note that upon rescaling (S1), only the quadratic part of the action remains temperature-dependent, with bare propagator

$$\hat{G}_0^T(\mathbf{k}, \nu_n) = T^{\frac{1}{2}} \left[ i\nu_n - \hat{\mathcal{H}}_{\mathbf{k}} \right]^{-1}. \tag{S2}$$

The temperature therefore takes the role of a renormalization group scale and one can derive an exact evolution equation for the effective action functional [33, 69]

$$\frac{d}{dT} \Gamma^T[\eta, \overline{\eta}] = -\sum_n \int_{\mathbf{k}} \overline{\eta}_{\mathbf{k},n} \frac{d}{dT} [\hat{G}_0^T(\mathbf{k}, \nu_n)]^{-1} \eta_{\mathbf{k},n} - \frac{1}{2} \text{Tr} \left[ \frac{d}{dT} [\mathbf{G}_0^T]^{-1} \left( \mathbf{\Gamma}^{(2)T}[\eta, \overline{\eta}] \right)^{-1} \right]. \tag{S3}$$

Here, we have defined

$$\mathbf{G}_0^T(\mathbf{k}, \nu_n) = \begin{pmatrix} \hat{G}_0^T(\mathbf{k}, \nu_n) & 0 \\ 0 & -\left[ \hat{G}_0^T(-\mathbf{k}, -\nu_n) \right]^T \end{pmatrix}, \tag{S4}$$

and

$$\mathbf{\Gamma}^{(2)T}[\eta, \overline{\eta}](x, x') = \begin{pmatrix} \frac{\delta \Gamma^T[\eta, \overline{\eta}]}{\delta \overline{\eta}(x') \delta \eta(x)} & \frac{\delta \Gamma^T[\eta, \overline{\eta}]}{\delta \overline{\eta}(x') \delta \eta(x)} \\ \frac{\delta \Gamma^T[\eta, \overline{\eta}]}{\eta(x') \delta \eta(x)} & \frac{\delta \Gamma^T[\eta, \overline{\eta}]}{\delta \eta(x') \delta \eta(x)} \end{pmatrix}, \tag{S5}$$

where $x = (\mathbf{k}, n, \sigma)$ and the symbol $[\bullet]^{\mathrm{T}}$ denotes matrix transposition and does not have to be confused with the temperature $T$. Expanding the effective action functional $\Gamma^T[\eta, \overline{\eta}]$ on both sides of (S3), one can derive an infinite hierarchy of flow equations for the $n$ particle vertices

$$\Gamma^{(2n)T}(x_1, \ldots, x_n, x_1', \ldots, x_n') = \frac{\delta \Gamma^T[\eta, \overline{\eta}]}{\delta \overline{\eta}(x_1') \ldots \delta \overline{\eta}(x_n') \delta \eta(x_1) \ldots \delta \eta(x_n)} \bigg|_{\eta, \overline{\eta} = 0}. \tag{S6}$$

Eq. (S3) is completed with an initial condition for the effective action reading as [33]

$$\Gamma^{T \to +\infty}[\eta, \overline{\eta}] = \mathcal{S}^{T \to +\infty}[\eta, \overline{\eta}]. \tag{S7}$$

In this work, we *truncate* the effective action at the two-particle level, that is, we set to zero all the three- and more particle correlators. We also ignore the flow of the self-energy, that is, we keep the one-particle vertex fixed to $(\hat{G}_0^T)^{-1}$.

The only vertex function we retain is the two particle vertex

$$V^{\sigma_1 \sigma_2 \sigma_3 \sigma_4, T}(\mathbf{k}_1, \mathbf{k}_2, \mathbf{k}_3, n_1, n_2, n_3) = \Gamma^{(4)T}. \tag{S8}$$

which depends, in principle, on the four spins of the two incoming and two outgoing electrons, and on three momenta and Matsubara frequencies. We introduce a further approximation by neglecting the frequency dependence.

To disentangle the dependence of $V^T$ on the three momenta, we employ a channel decomposition [70, 71]:

$$V^{\sigma_1 \sigma_2 \sigma_3 \sigma_4}(\mathbf{k}_1, \mathbf{k}_2, \mathbf{k}_3) = U\left(\delta_{\sigma_1 \sigma_3} \delta_{\sigma_2 \sigma_4} - \delta_{\sigma_1 \sigma_4} \delta_{\sigma_2 \sigma_3}\right) \\ + P^{\sigma_1 \sigma_2 \sigma_3 \sigma_4}_{\mathbf{k}_1, \mathbf{k}_3}(\mathbf{k}_1 + \mathbf{k}_2) + D^{\sigma_1 \sigma_2 \sigma_3 \sigma_4}_{\mathbf{k}_1, \mathbf{k}_4}(\mathbf{k}_3 - \mathbf{k}_1) - D^{\sigma_1 \sigma_2 \sigma_4 \sigma_3}_{\mathbf{k}_1, \mathbf{k}_3}(\mathbf{k}_2 - \mathbf{k}_3), \tag{S9}$$

where, to simplify the notation, we have dropped the $T$-dependence. We refer to $P$ and $D$ as particle-particle and particle-hole channels, respectively. It is clear from Eq. (S7) that at $T \to +\infty$ one has $V = U$ and therefore $P = D = 0$. The flow equations read as

$$\frac{d}{dT} P^{\sigma_1 \sigma_2 \sigma_3 \sigma_4}_{\mathbf{k}, \mathbf{k}'}(\mathbf{q}) = -\sum_{\substack{s_1, s_2 \\ s_3, s_4}} \int_{\mathbf{p}} V^{\sigma_1 \sigma_2 s_3 s_4}(\mathbf{k}, \mathbf{q} - \mathbf{k}, \mathbf{p}) \, \dot{\Pi}^{s_1 s_2 s_3 s_4}_{P, \mathbf{p}}(\mathbf{q}) \, V^{s_1 s_2 \sigma_3 \sigma_4}(\mathbf{p}, \mathbf{q} - \mathbf{p}, \mathbf{k}'), \tag{S10a}$$

$$\frac{d}{dT} D^{\sigma_1 \sigma_2 \sigma_3 \sigma_4}_{\mathbf{k}, \mathbf{k}'}(\mathbf{q}) = 2\sum_{\substack{s_1, s_2 \\ s_3, s_4}} \int_{\mathbf{p}} V^{\sigma_1 s_2 \sigma_3 s_4}(\mathbf{k}, \mathbf{p} + \mathbf{q}, \mathbf{k} + \mathbf{q}) \, \dot{\Pi}^{s_1 s_2 s_3 s_4}_{D, \mathbf{p}}(\mathbf{q}) \, V^{s_1 \sigma_2 s_3 \sigma_4}(\mathbf{p}, \mathbf{k}' + \mathbf{q}, \mathbf{p} + \mathbf{q}), \tag{S10b}$$

with

$$\Pi^{s_1 s_2 s_3 s_4}_{P, \mathbf{p}}(\mathbf{q}) = \sum_n \left[G^T_{0, s_3 s_1}(\mathbf{p}, \nu_n) G^T_{0, s_4 s_2}(\mathbf{q} - \mathbf{p}, -\nu_n)\right], \tag{S11a}$$

$$\Pi^{s_1 s_2 s_3 s_4}_{D, \mathbf{p}}(\mathbf{q}) = \sum_n \left[G^T_{0, s_4 s_1}(\mathbf{p}, \nu_n) G^T_{0, s_3 s_2}(\mathbf{p} + \mathbf{q}, \nu_n)\right]. \tag{S11b}$$

The symbol $\dot{\Pi}$ is a shorthand for $d\Pi/dT$. To simplify the treatment, we parametrize the spin dependence of $P$ and $D$ as

$$P^{\sigma_1 \sigma_2 \sigma_3 \sigma_4}_{\mathbf{k}, \mathbf{k}'}(\mathbf{q}) = \frac{1}{2} \sum_{\alpha, \beta = 0}^{3} P^{\alpha\beta}_{\mathbf{k}, \mathbf{k}'}(\mathbf{q}) \, t^{\alpha}_{\sigma_1 \sigma_2} \left(t^{\beta}_{\sigma_3 \sigma_4}\right)^{\dagger}, \tag{S12a}$$

$$D^{\sigma_1 \sigma_2 \sigma_3 \sigma_4}_{\mathbf{k}, \mathbf{k}'}(\mathbf{q}) = \frac{1}{2} \sum_{\alpha, \beta = 0}^{3} D^{\alpha\beta}_{\mathbf{k}, \mathbf{k}'}(\mathbf{q}) \, \tau^{\alpha}_{\sigma_1 \sigma_3} \tau^{\beta}_{\sigma_2 \sigma_4}, \tag{S12b}$$

where $\tau^{1,2,3}$ are the Pauli matrices, $\tau^0 = \mathbb{1}$, and $t^{\alpha} = i\tau^{\alpha}\tau^2$. To treat the dependence of the channels on $\mathbf{k}$ and $\mathbf{k}'$, we resort to a *truncated unity* approach [72, 73], that is, we expand them in a complete basis of form factors $\{f^{\ell}_{\mathbf{k}}\}$:

$$X^{\alpha\beta}_{\mathbf{k}, \mathbf{k}'}(\mathbf{q}) = \sum_{\ell, \ell'} X^{\alpha\beta}_{\ell\ell'}(\mathbf{q}) f^{\ell}_{\mathbf{k}} (f^{\ell'}_{\mathbf{k}'})^*, \tag{S13}$$

with $X = P$ or $D$, and truncate the sum up to a given order. We choose form factors that transform in the irreducible representations of the lattice symmetry group $C_{4v}$. It is possible to choose $f^{\ell}_{\mathbf{k}}$ in such a way that their real space

representation takes nonzero values only in a given shell of neighbors of a given reference site [74]. It is therefore convenient to truncate the sum (S13) by including all form factors up to a given shell. For our calculations, we truncate the sum to the first shell of neighbors, that is, we only consider the form factors

$$f_{\mathbf{k}}^0 = 1, \tag{S14a}$$

$$f_{\mathbf{k}}^1 = \cos k_x + \cos k_y, \tag{S14b}$$

$$f_{\mathbf{k}}^2 = \cos k_x - \cos k_y, \tag{S14c}$$

$$f_{\mathbf{k}}^3 = \sqrt{2} \sin k_x, \tag{S14d}$$

$$f_{\mathbf{k}}^4 = \sqrt{2} \sin k_y. \tag{S14e}$$

Inserting the identity

$$\delta_{\mathbf{k},\mathbf{k}'} = \sum_\ell f_{\mathbf{k}}^\ell (f_{\mathbf{k}'}^\ell)^* \tag{S15}$$

into (S10), we obtain

$$\frac{d}{dT} X_{\ell\ell'}^{\alpha\beta}(\mathbf{q}) = \frac{1}{2} \zeta_X \sum_{\alpha',\beta'} \sum_{m,m'} X[V]_{\ell m}^{\alpha\beta'}(\mathbf{q}) \, \dot{\Pi}_{X,mm'}^{\beta'\alpha'}(\mathbf{q}) \, X[V]_{m'\ell'}^{\alpha'\beta}(\mathbf{q}), \tag{S16}$$

with $X = P$ or $D$, $\zeta_P = -1$ and $\zeta_D = +2$. The symbols $P[V]$ and $D[V]$ represent the projection of the full vertex (S9) onto the $P$ and $D$ channels and they read as

$$P[V]_{\ell\ell'}^{\alpha\beta}(\mathbf{q}) = \int_{\mathbf{k},\mathbf{k}'} \frac{1}{2} \sum_{\substack{\sigma_1,\sigma_2 \\ \sigma_3,\sigma_4}} V^{\sigma_1\sigma_2\sigma_3\sigma_4}(\mathbf{k}, \mathbf{q} - \mathbf{k}, \mathbf{k}')(t_{\sigma_1\sigma_2}^\alpha)^\dagger t_{\sigma_3\sigma_4}^\beta (f_{\mathbf{k}}^\ell)^* f_{\mathbf{k}'}^{\ell'}, \tag{S17a}$$

$$D[V]_{\ell\ell'}^{\alpha\beta}(\mathbf{q}) = \int_{\mathbf{k},\mathbf{k}'} \frac{1}{2} \sum_{\substack{\sigma_1,\sigma_2 \\ \sigma_3,\sigma_4}} V^{\sigma_1\sigma_2\sigma_3\sigma_4}(\mathbf{k}, \mathbf{k}' + \mathbf{q}, \mathbf{k}) \tau_{\sigma_3\sigma_1}^\alpha \tau_{\sigma_4\sigma_2}^\beta (f_{\mathbf{k}}^\ell)^* f_{\mathbf{k}'}^{\ell'}. \tag{S17b}$$

Similar expressions hold for the bubbles $\dot{\Pi}_{X,\ell\ell'}^{\beta\alpha}(\mathbf{q})$. For more details on how to perform the vertex projections, see for example Refs. [72, 73].

We run a flow starting from a high-temperature value $T_{\text{ini}} \simeq 10t$ at which $P$ and $D$ can be initialized by a second order perturbation theory result:

$$X_{\ell\ell'}^{\alpha\beta\,T_{\text{ini}}}(\mathbf{q}) = \frac{1}{2} \zeta_X U^2 \Pi_{X,00}^{\alpha\beta\,T_{\text{ini}}}(\mathbf{q}) \delta_{\ell,0} \delta_{\ell',0}, \tag{S18}$$

to (approximately) account for the integration from $T = +\infty$ to $T = T_{\text{ini}}$. We stop the flow at a temperature $T_c$ where one of $P$ or $D$ exceeds the value $25t$ or at a temperature $T_{\text{min}} = 10^{-2}t$ where the momentum integration in the projections of the bubbles (S11) onto form factors becomes numerically unstable.

## B. Renormalized mean-field theory

To continue the flow below the stopping temperature $T_s \equiv \max(T_c, T_{\text{min}})$, and therefore detect the leading superconducting instabilities, we resort to a combination of mean-field theory with the functional renormalization group [34, 35, 37]. This approximation accounts for continuing the flow in the superconducting regime, that is, by introducing a pairing gap and anomalous vertices, by keeping only particle-particle diagrams. The simplified flow equations can be formally integrated and cast in the form of *renormalized* mean-field equations. In this way, one obtains the gap equation (valid for any $0 \leq T \leq T_s$)

$$\Delta_{\mathbf{k}}^{\sigma\sigma'} = \sum_{s,s'} \int_{\mathbf{k}'} \widetilde{V}_{\mathbf{k}\mathbf{k}'}^{\sigma\sigma's s'} \left[ T \sum_n F^{ss'}(\mathbf{k}', \nu_n) \right],$$

where $F^{\sigma\sigma'}(\mathbf{k}, \nu_n)$ are the spin components of the anomalous propagator $\hat{F}(\nu_n)$, defined through

$$\mathbf{G}_{\text{BdG}}(\mathbf{k}, i\nu_n) = \begin{pmatrix} \hat{G}(\mathbf{k}, \nu_n) & \hat{F}(\mathbf{k}, \nu_n) \\ \hat{F}^\dagger(\mathbf{k}, \nu_n) & -[\hat{G}(-\mathbf{k}, -\nu_n)]^{\mathrm{T}} \end{pmatrix} = \begin{pmatrix} i\nu_n - \hat{\mathcal{H}}_{\mathbf{k}} & -\hat{\Delta}_{\mathbf{k}} \\ -\hat{\Delta}_{\mathbf{k}}^\dagger & i\nu_n + [\hat{\mathcal{H}}_{-\mathbf{k}}]^{\mathrm{T}} \end{pmatrix}^{-1}, \tag{S19}$$

with $\hat{\Delta}_{\mathbf{k}}$ the gap matrix (in spin space). The function $\widetilde{V}_{\mathbf{k}\mathbf{k}'}^{\sigma_1\sigma_2\sigma_3\sigma_4}$ is the two-particle-irreducible vertex function in the particle-particle channel at zero center-of-mass momentum. It is determined by inverting a Bethe-Salpeter equation at the stopping temperature $T_s$:

$$V^{\sigma_1\sigma_2\sigma_3\sigma_4,T_s}(\mathbf{k},-\mathbf{k},\mathbf{k}') = \widetilde{V}_{\mathbf{k}\mathbf{k}'}^{\sigma_1\sigma_2\sigma_3\sigma_4} + \sum_{\substack{s_1,s_2\\s_3,s_4}}\int_{\mathbf{p}}\widetilde{V}_{\mathbf{k}\mathbf{p}}^{\sigma_1\sigma_2 s_3 s_4}\,\Pi_{P,\mathbf{p}}^{s_1 s_2 s_3 s_4,T_s}(\mathbf{q}=\mathbf{0})\,V^{s_1 s_2 \sigma_3\sigma_4,T_s}(\mathbf{p},-\mathbf{p},\mathbf{k}'). \tag{S20}$$

Eq. (S20) can be inverted with an approach similar to the one presented in the previous section: one can paramterize the spin dependence of $\widetilde{V}$, $V^{T_s}$, and $\Pi_P^{T_s}$ by means of the $t^\alpha$ matrices (see Eq. (S12a)) and expand the $\mathbf{k}$- and $\mathbf{k}'$-dependence in form factors and truncate the expansion at a given order. Notice that the order kept in this inversion can be also higher than the one used in the fRG calculation of $V^{T_s}$. One obtains

$$P[V^{T_s}]_{\ell\ell'}^{\alpha\beta}(\mathbf{q}=\mathbf{0}) = 2\widetilde{V}_{\ell\ell'}^{\alpha\beta} + \sum_{\alpha',\beta'}\sum_{m,m'}\widetilde{V}_{\ell m}^{\alpha\beta'}\,\Pi_{P,mm'}^{\beta'\alpha',T_s}(\mathbf{q}=\mathbf{0})\,P[V^{T_s}]_{m'\ell'}^{\alpha'\beta}(\mathbf{q}=\mathbf{0}). \tag{S21}$$

Defining a multi-index $\mathbf{i} = (\alpha,\ell)$, $\widetilde{V}$, $P[V^{T_s}]$, and $\Pi_P^{T_s}$ become matrices, and $\widetilde{V}$ can be cast as

$$\widetilde{V}_{\mathbf{i}\mathbf{i}'} = \left\{\left[[2\mathbf{P}[V^{T_s}](\mathbf{q}=\mathbf{0})]^{-1} - \mathbf{\Pi}_P^{T_s}(\mathbf{q}=\mathbf{0})\right]^{-1}\right\}_{\mathbf{i}\mathbf{i}'}. \tag{S22}$$

By expanding also the gap in form factors and $t^\alpha$ matrices, the gap equation takes the form

$$\Delta_\ell^\alpha = \sum_{\beta,\ell'}\widetilde{V}_{\ell\ell'}^{\alpha\beta}\,\mathcal{F}_{\ell'}^\beta, \tag{S23}$$

where we have defined

$$\Delta_{\mathbf{k}}^{\sigma\sigma'} = \sum_{\alpha=0}^3\sum_\ell \Delta_\ell^\alpha t_{\sigma\sigma'}^\alpha f_{\mathbf{k}}^\ell, \tag{S24}$$

and

$$\mathcal{F}_{\ell'}^\beta = \frac{1}{2}T\sum_n\int_{\mathbf{k}}\text{Tr}\left[F(\mathbf{k},\nu_n)(t^\alpha)^\dagger\right](f_{\mathbf{k}}^\ell)^*. \tag{S25}$$

Notice that Eq. (S23) still depends on the temperature via $\mathcal{F}_\ell^\alpha$.

Eq. (S23) can be used to determine the gap function at a given temperature $T \geq T_s$ but also to simply determine the leading superconducting state. This can be achieved by *linearizing* it with respect to $\Delta_\ell^\alpha$. By noticing that

$$\left.\frac{\delta\mathcal{F}_\ell^\alpha}{\delta\Delta_{\ell'}^\beta}\right|_{\Delta_\ell^\alpha=0} = \Pi_{P,\ell\ell'}^{\alpha\beta}(\mathbf{q}=\mathbf{0}), \tag{S26}$$

we obtain

$$\Delta_\ell^\alpha = \sum_{\beta,\gamma}\sum_{\ell',m}\widetilde{V}_{\ell\ell'}^{\alpha\beta}\Pi_{P,\ell'm}^{\beta\gamma}(\mathbf{q}=\mathbf{0})\Delta_m^\gamma. \tag{S27}$$

For temperatures close to the superconducting transition temperature $T_{\text{SC}}$, where the overall magnitude of the gap is small and the linearization carried out above is justifiable, the gap is an eigenvector of the matrix $\widetilde{V}\circ\Pi_P(\mathbf{q}=\mathbf{0})$ with eigenvalue 1. $T_{\text{SC}}$ can therefore be determined as the temperature at which the largest positive eigenvalue of the matrix $\widetilde{V}\circ\Pi_P(\mathbf{q}=\mathbf{0})$ becomes 1. Since in a Fermi liquid some components of $\Pi_P(\mathbf{q}=\mathbf{0})$ diverge like $\log(T^{-1})$ (or $\log^2(T^{-1})$ at a van Hove singularity) as $T$ approaches 0, one will always find a finite value of $T_{\text{SC}}$, or, in other words, Eq. (S23) has always a nontrivial solution at $T=0$. The numerical values of $\Delta_\ell^\alpha$, however, can be exponentially small, so that their calculation becomes numerically involved. For this reason, we analyze the linearized gap equation, Eq. (S27), at a low temperature, $T_{\text{low}}=10^{-6}t$, and determine the leading superconducting eigenvalue $\lambda_{\text{SC}}$, defined as the largest positive eigenvalue of $\widetilde{V}\circ\Pi_P(\mathbf{q}=\mathbf{0})$. The symmetry of the corresponding eigenvector determines the symmetry of the leading superconducting instability.

|  | $\mathbf{k}$ | $\vec{\tau}$ | $\hat{U}$ |
|---|---|---|---|
| E | $k_x \to k_x$ <br> $k_y \to k_y$ | $\tau^1 \to \tau^1$ <br> $\tau^2 \to \tau^2$ <br> $\tau^3 \to \tau^3$ | $\mathbb{1}$ |
| $C_4^{\pm}$ | $k_x \to \mp k_y$ <br> $k_y \to \pm k_x$ | $\tau^1 \to \mp\tau^2$ <br> $\tau^2 \to \pm\tau^1$ <br> $\tau^3 \to \tau^3$ | $e^{\pm i \frac{\pi}{4}\tau^3}$ |
| $C_2$ | $k_x \to -k_x$ <br> $k_y \to -k_y$ | $\tau^1 \to -\tau^1$ <br> $\tau^2 \to -\tau^2$ <br> $\tau^3 \to \tau^3$ | $e^{i\frac{\pi}{2}\tau^3}$ |
| $\Sigma_v^{x(y)}$ | $k_x \to \pm k_x$ <br> $k_y \to \mp k_y$ | $\tau^1 \to \mp\tau^1$ <br> $\tau^2 \to \pm\tau^2$ <br> $\tau^3 \to \tau^3$ | $i\tau^3\mathcal{K}$ ($\mathbb{1}\mathcal{K}$) |
| $\Sigma_d^{1(2)}$ | $k_x \to \pm k_y$ <br> $k_y \to \pm k_x$ | $\tau^1 \to \mp\tau^2$ <br> $\tau^2 \to \mp\tau^1$ <br> $\tau^3 \to \tau^3$ | $e^{\pm i\frac{\pi}{4}\tau^3}\mathcal{K}$ |

Table SI. Discrete spin-lattice transformations that leave the Rashba-Hubbard Hamiltonian invariant. In the first column we show their effect on the lattice momentum $\mathbf{k}$, in the second one how they transform the Pauli matrices $\vec{\tau}$, and in the third one the operator that implements the spin part of the transformation on the electron operators. We remark that the transformation of the Pauli matrices is implemented as $\vec{\tau} \to \hat{U}^\dagger \vec{\tau} \hat{U}$. By $\mathcal{K}$ we denote the complex conjugation operator.

|  | E | $C_4^{\pm}$ | $C_2$ | $\Sigma_v^{x(y)}$ | $\Sigma_d^{1(2)}$ | $\hat{\Delta}_{\mathbf{k}}$ |
|---|---|---|---|---|---|---|
| $A_1$ | 1 | 1 | 1 | 1 | 1 | $g_{\mathbf{k}}^{A_1} t^0 - g_{\mathbf{k}}^{E,y} t^1 + g_{\mathbf{k}}^{E,x} t^2$ |
| $A_2$ | 1 | 1 | 1 | -1 | -1 | $g_{\mathbf{k}}^{A_2} t^0 + g_{\mathbf{k}}^{E,x} t^1 + g_{\mathbf{k}}^{E,y} t^2$ |
| $B_1$ | 1 | -1 | 1 | 1 | -1 | $g_{\mathbf{k}}^{B_1} t^0 + g_{\mathbf{k}}^{E,y} t^1 + g_{\mathbf{k}}^{E,x} t^2$ |
| $B_2$ | 1 | -1 | 1 | -1 | 1 | $g_{\mathbf{k}}^{B_2} t^0 + g_{\mathbf{k}}^{E,x} t^1 - g_{\mathbf{k}}^{E,y} t^2$ |
| $E$ | $\begin{pmatrix} 1 & 0 \\ 0 & 1 \end{pmatrix}$ | $\begin{pmatrix} 0 & \pm 1 \\ \mp 1 & 0 \end{pmatrix}$ | $\begin{pmatrix} -1 & 0 \\ 0 & -1 \end{pmatrix}$ | $\begin{pmatrix} \pm 1 & 0 \\ 0 & \mp 1 \end{pmatrix}$ | $\begin{pmatrix} 0 & \pm 1 \\ \pm 1 & 0 \end{pmatrix}$ | $\left( g_{\mathbf{k}}^{E,x} t^3, g_{\mathbf{k}}^{E,y} t^3 \right)$ |

Table SII. Transformation properties of the irreps under the action of the spin-lattice symmetry group. In the two-dimensional irrep $E$ the transformation matrices depend on the basis choice. In the last column, we show the form of the gap function in each irrep. The functions $g_{\mathbf{k}}^{A_1}$, $g_{\mathbf{k}}^{A_2}$, $g_{\mathbf{k}}^{B_1}$ and $g_{\mathbf{k}}^{B_2}$ are irreps of the sole $C_{4v}$, while $g_{\mathbf{k}}^{E,x}$ and $g_{\mathbf{k}}^{E,y}$ are a basis of the two-dimensional irrep $E$ of $C_{4v}$ chosen such that they transform under $C_{4v}$ as shown in the last line of the table. The lowest harmonic contributions for these functions are listed in Table SIII.

## SII. SYMMETRY CLASSIFICATION OF THE SUPERCONDUCTING STATES

The symmetry classification of the superconducting symmetries relies on the symmetry group of the Hamiltonian (1). Because of the Rashba spin-orbit coupling, the lattice point group $C_{4v}$ does not leave the Hamiltonian invariant. We have therefore to consider a different group that combines $C_{4v}$ with spin rotations. This group can be simply obtained from the square lattice point group by adding to each of its elements a spin transformation that leaves the Rashba term $\sim (-\sin k_y \, \tau^1 + \sin k_x \, \tau^2)$ invariant. In Table SI we list all the discrete combined spin-lattice transformations that leave the Hamiltonian (1) invariant. These symmetries form a discrete group with eight elements $\{\text{E}, \text{C}_4^+, \text{C}_4^-, \text{C}_2, \Sigma_v^x, \Sigma_v^y, \Sigma_d^1, \Sigma_d^2\}$. Since this group is equivalent to the point group $C_{4v}$, its irreducible representations (irreps) are the same: four are one-dimensional ($A_1$, $A_2$, $B_1$, $B_2$), and one two-dimensional ($E$). We can therefore classify the symmetry of the superconducting gap according to these irreps. We decompose the gap into form factors

| $g_{\mathbf{k}}^{A_1}$ | $1,\ \cos k_x + \cos k_y,\ 2\cos k_x \cos k_y,\ \cos(2k_x)+\cos(2k_y),\ \ldots$ |
|---|---|
| $g_{\mathbf{k}}^{A_2}$ | $2\sqrt{2}\,(\cos k_x - \cos k_y)\sin k_x \sin k_y,\ \ldots$ |
| $g_{\mathbf{k}}^{B_1}$ | $\cos k_x - \cos k_y,\ \cos(2k_x)-\cos(2k_y),\ \ldots$ |
| $g_{\mathbf{k}}^{B_2}$ | $2\sin k_x \sin k_y,\ 2\sqrt{2}\,(\cos k_x + \cos k_y)\sin k_x \sin k_y,\ \ldots$ |
| $\left(g_{\mathbf{k}}^{E,x},g_{\mathbf{k}}^{E,y}\right)$ | $\left(\sqrt{2}\sin k_x,\sqrt{2}\sin k_y\right),\ (2\cos k_y \sin k_x, 2\cos k_x \sin k_y),\ \ldots$ |

Table SIII. Lowest harmonics for the functions $g_{\mathbf{k}}^{A_1}$, $g_{\mathbf{k}}^{A_2}$, $g_{\mathbf{k}}^{B_1}$, $g_{\mathbf{k}}^{B_2}$, $g_{\mathbf{k}}^{E,x}$, $g_{\mathbf{k}}^{E,y}$ shown in Table SII. Note that we have employed the normalization convention $\int_{\mathbf{k}} |g_{\mathbf{k}}^{\chi}|^2 = 1$.

and $t^\alpha$ matrices, according to Eq. (S24). We notice that the action of the group on the matrices $t^\alpha$ is given by

$$t^\alpha \to \hat{U}^\dagger t^\alpha \hat{U}^*,$$ 
(S28)

with the different transformation matrices $\hat{U}$ listed in Table SI. As a consequence, when $\alpha = 1, 2, 3$, the matrices $t^\alpha$ transform exactly as the Pauli matrices (as shown in Table SI), while $t^0$ transforms always trivially. We also remark that, in order to preserve the overall antisymmetry of the pairing wavefunction, the singlet (triplet) components of the gap function $\Delta_\ell^\alpha$, corresponding to $\alpha = 0$ ($\alpha = 1, 2, 3$), must be symmetric (antisymmetric) for $\mathbf{k} \to -\mathbf{k}$. For this reason, only the pairs of indices $(\alpha, \ell)$ with $\alpha = 0$ and $f_{\mathbf{k}}^\ell$ even or $\alpha = 1, 2, 3$ and $f_{\mathbf{k}}^\ell$ odd are allowed in $\Delta_\ell^\alpha$. In Table SII, we list the irreps of the combined lattice-spin discrete symmetry group, together with the typical form that the gap function $\hat{\Delta}_{\mathbf{k}}$ takes in each one of them. In Table SIII we show the lowest harmonics for the functional dependence of $\hat{\Delta}_{\mathbf{k}}$ on $\mathbf{k}$. Note that equal spin triplet pairing (described with the matrices $t^1$ and $t^2$) can only occur in combination with singlet pairing (described by $t^0$) in a one-dimensional representation. On the other hand, opposite spin triplet pairing ($t^3$) can take place only in the two-dimensional irrep $E$.

## SIII. REAL SPACE RENORMALIZED MEAN-FIELD

In Sec. SI B, we solve the renormalized mean-field Hamiltonian in momentum space due to a translational invariance. In this section, we give the details of solving the renormalized mean-field Hamiltonian in real space. In the absence of any inhomogeneities, the two approaches are analogues and give the same results. However, inhomoegeneities arising either due to edges/surfaces or disorder can only be captured in a real space picture. Here, we ignore any disorder and focus on the edge properties and possible symmetry breakings within the real space picture. We decouple $H_U$ in Eq. (1) only in the Cooper channel and the resultant real space Hamiltonian is,

$$H_{\text{eff}} = H_0 + \sum_{\substack{i\mu \\ \sigma\sigma'}} \Delta_\mu^{\sigma\sigma'}(i)c_{i+\mu,\sigma'}^\dagger c_{i,\sigma}^\dagger + \text{H.c.},$$ 
(S29)

where $\mu$ denotes the nearest neighbor bonds $\pm x, \pm y$ of the lattice site $i$, $H_0$ is defined in Eq. (1) and $\Delta_\mu^{\sigma\sigma'}$ is the superconducting order parameter given by

$$\Delta_\mu^{\sigma\sigma'}(i) = \sum_{s,s'} \sum_\nu \widetilde{V}_{\mu\nu}^{\sigma\sigma'ss'} \langle c_{i,s} c_{i+\nu,s'} \rangle,$$ 
(S30)

where the interaction strengths are

$$\widetilde{V}_{\mu\nu}^{\sigma\sigma'ss'} = \frac{1}{2} \sum_{\alpha\beta} \sum_{\ell\ell'} \widetilde{V}_{\ell\ell'}^{\alpha\beta} t_{\sigma\sigma'}^\alpha (t_{ss'}^\beta)^\dagger f^\ell(\mu) [f^{\ell'}(\nu)]^*,$$ 
(S31)

with $f^\ell(\mu)$ being the Fourier transform of $f_{\mathbf{k}}^\ell$ and $\widetilde{V}_{\ell\ell'}^{\alpha\beta}$ are obtained from the fRG calculation. $H_{\text{MF}}$ is diagonlized using the Bogoliubov-de Gennes transformations [75], $c_{i\sigma} = \sum_n (\gamma_n u_{i\sigma}^n - \sigma\gamma_n^\dagger v_{i\sigma}^{n\,*})$, where $\gamma_{n\sigma}^\dagger$ ($\gamma_{n\sigma}$) are the creation

(annihilation) operators of the Bogoliubov quasiparticle at a state $n$, and $u_{i\sigma}^n$ and $v_{i\sigma}^n$ the eigenfunctions with eigenvalues $E$. The resulting eigen-system is then solved self-consistently for the superconducting order parameters $\Delta_\mu^{\sigma\sigma'}$. Moreover, the findings of the order paramters obtained in the momentum space suggest that the dominant pairing is only on nearest neighbors, i.e., either $A_1$ or $B_1$ phases, for most parts of the phase diagram except in a narrow region of the phase diagram ($B_2$ phase red region in Fig. 1). Hence, we restrict the order parameters only to nearest neighbors in real space, as already expressed in Eq. (S29). This helps to reduce the numer of independent self-consistent order parameters to 16 per lattice site. We then project local $\Delta_\mu^{\sigma\sigma'}$ on onsite equivalents of different pairing components to analyse the symmetry of the order parameter. As shown in Tables SII and SIII, the only possible nearest neighbour singlet components are $d$-wave and extended $s$-wave order parameters. The onsite equivalents of the singlets are given as

$$\Delta_{d/s}(i) = \frac{1}{8} \sum_\mu \epsilon_\mu^{d/s} \sum_\sigma \text{sgn}(\sigma) \Delta_\mu^{\sigma\bar{\sigma}}(i),$$ (S32)

where $\Delta_d$ is the $d$-wave order parameter, which in momentum space is proportional to $(\cos k_x - \cos k_y)$. $\Delta_s$ is the extended $s$-wave order parameter, which in momentum space is proportional to $(\cos k_x + \cos k_y)$. The perfactor $\epsilon_\mu^{d/s}$ is $\epsilon_\mu^{d/s} = 1$ for $\mu = \pm x$ and $\epsilon_\mu^{d/s} = -1/1$ for $\mu = \pm y$. The phase of the $d$-wave order parameter is characterized with $\theta$ with the definition $\Delta_d(i) = |\Delta_d(i)| \exp(i\theta)$. The triplet order parameters are mainly $p$-wave with the onsite equivalent given as

$$\Delta_{p_\nu}^{\sigma\sigma'}(i) = \frac{1}{2i} \sum_\mu \text{sgn}(\mu) \Delta_\mu^{\sigma\sigma'},$$ (S33)

where $\Delta_{p_x/p_y}^{\sigma\sigma'}$ are the $p$-wave order parameters, which in momentum space is proportional to $\sin k_x$ or $\sin k_y$. Different spin components of the $p$-wave order parameters can be combined to obtain the order parameters in the irreducible represenations as

$$\Delta_{A_1}^t(i) = \frac{1}{4} \left( \sum_\sigma \text{sgn}(\sigma) \Delta_{p_y}^{\sigma\sigma} - i \sum_\sigma \Delta_{p_x}^{\sigma\sigma} \right),$$ (S34)

$$\Delta_{B_1}^t(i) = \frac{1}{4} \left( \sum_\sigma \text{sgn}(\sigma) \Delta_{p_y}^{\sigma\sigma} + i \sum_\sigma \Delta_{p_x}^{\sigma\sigma} \right).$$ (S35)

Other mixed-spin triplet components are negligible compared to the equal spin triplets.

The interaction strengths $\tilde{V}_{\mu\nu}^{\sigma\sigma'ss'}$ obtained from fRG give rise to very small values of the superconducting order parameters in most parts of the phase diagram. As a result, the corresponding coherence lengths are large giving the necessity of working with extremely large system sizes to avoid interference effects of the edges while studying the topological edge states. Hence, we scale the interaction strengths obtained from fRG by 10 in order to solve the gap equation Eq. (S30) in real space for all representative points in the phase diagram investigated. We have verified that the qualitative features do not change with reducing the scale factor. We first match the self-consistent superconducting order parameters with periodic boundary condition using the solutions of the momentum space and real space gap equations. We then use open boundary conditions to investigate each of the phases. The choice of the geometry of the boundary depends on the characteristic edge states of different superconducting phases, as explained in the main text.

## SIV. EFFECTIVE MODEL AROUND VAN HOVE SINGULARITY POINTS

In this section we derive low-energy Hamiltonians near the two VHS points of the square lattice and discuss the logarithmic divergences in the DOS near these VHSs. The non-interacting Hamiltonian matrix of the Rashba model on the square lattice is given by

$$h(\mathbf{k}) = (\epsilon_\mathbf{k} - \mu)\tau_0 + \lambda \vec{g}_\mathbf{k} \cdot \vec{\tau},$$ (S36)

where $\epsilon_\mathbf{k} = -2t(\cos k_x + \cos k_y) - 4t' \cos k_x \cos k_y$ and $\vec{g}_\mathbf{k} = 2t(-\sin k_y, \sin k_x, 0)$. The eigenvalues are $E_\mathbf{k} = (\epsilon_\mathbf{k} - \mu) \pm 2t\lambda \sqrt{\sin^2 k_x + \sin^2 k_y}$. The chemical potentials of the two VHS points are $\mu_{\text{VHS}} = \pm 2(\sqrt{(t \pm 2t')^2 + t^2\lambda^2} - t)$. The

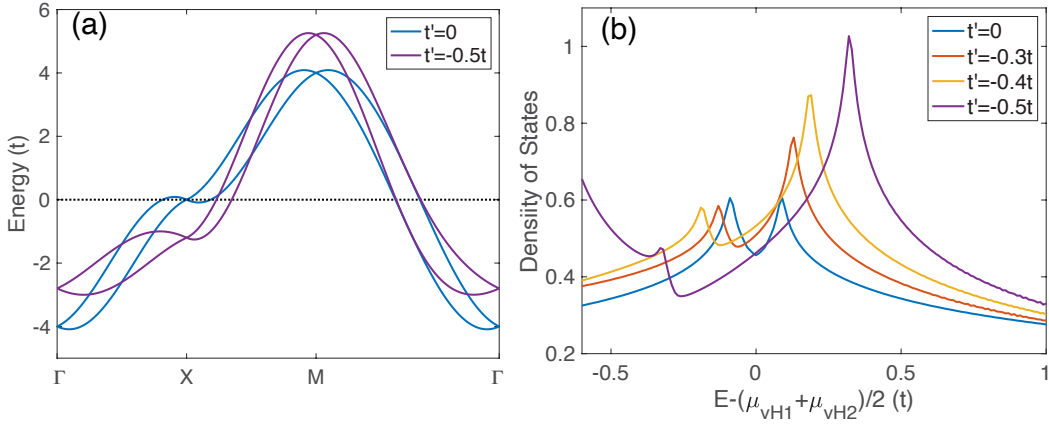

Figure S1. Band structures (a) and density of states (DOS) (b) around two the VHSs of the square lattice as a function of the second neighbor hopping $t'$.

low-energy Hamiltonian around the VHS point $(k_c, 0)$ on the $\Gamma - X$ direction (VHS1) reads

$$
\begin{aligned}
\epsilon(k_c + q_x, q_y) &= -2\left[(\cos k_c - \sin k_c \lambda + 1) + 2\cos k_c t'\right] + q_x^2\left[t(\cos k_c - \sin k_c \lambda) + 2\cos k_c t'\right] \\
&\quad + q_y^2\left(2\cos k_c t' + \frac{|\lambda|t}{\sin k_c} + t\right), \\
&= -2\left[t(\cos k_c - \sin k_c \lambda + 1) + 2\cos k_c t'\right] - q_x^2 \sin k_c \frac{(t + 2t')^2 + t^2\lambda^2}{\lambda t} + q_y^2 \frac{2\sin k_c \cos k_c t' + t\sin k_c + |\lambda|t}{\sin k_c},
\end{aligned}
\tag{S37}
$$

where $k_c$ is given by $\tan k_c = -\frac{\lambda t}{t + 2t'}$. The dispersion can be further simplified to

$$
\epsilon(k_c + q_x, q_y) = 2(D_1 - t) - q_x^2 D_1 + q_y^2\left[-\frac{2(t + 2t')t'}{D_1} + D_1 + t\right],
\tag{S38}
$$

with $D_1 = \sqrt{(t + 2t')^2 + \lambda^2 t^2}$.

The low-energy Hamiltonian around the VHS point $(\pi, k_c')$ on the $X - M$ direction (VHS2) reads

$$
\begin{aligned}
\epsilon(\pi + q_x, k_c' + q_y) &= 2\left[t(-\cos k_c' - \sin k_c \lambda + 1) + 2\cos k_c' t'\right] + q_y^2\left[t(\cos k_c' + \sin k_c' \lambda) - 2\cos k_c' t'\right] \\
&\quad - q_x^2\left(2\cos k_c' t' + \frac{|\lambda|t}{\sin k_c'} + t\right) \\
&= 2\left[t(-\cos k_c' - \sin k_c \lambda + 1) + 2\cos k_c' t'\right] + q_y^2 \sin k_c \frac{(t - 2t')^2 + t^2\lambda^2}{\lambda t} \\
&\quad - q_x^2\left(2\cos k_c' t' + \frac{|\lambda|t}{\sin k_c'} + t\right),
\end{aligned}
\tag{S39}
$$

where $k_c'$ is given by $\tan k_c' = \frac{\lambda t}{t - 2t'}$. The dispersion can be further simplified to

$$
\epsilon(\pi + q_x, q_y) = -2(D_2 - t) + q_y^2 D_2 - q_x^2\left[\frac{2(t - 2t')t'}{D_2} + D_2 + t\right],
\tag{S40}
$$

with $D_2 = \sqrt{(t - 2t')^2 + \lambda^2 t^2}$. The two VHSs are derived from the saddle at the X/Y point with Rashba spin-orbit coupling. In the absence of the next-nearest hopping $t'$, the dispersion around these two VHSs are similar and the corresponding effective masses along $x$ and $y$ are the same, leading to the same divergent density of states (see blue curves in Fig. S1). With increasing $|t'|$ (negative $t'$), the VHS1 on the $\Gamma - X$ line rapidly moves away from the X point, while the VHS2 on the $X - M$ line slightly move closer to the X point. As the coefficient of the logarithmically divergent DOS is proportional to $\sqrt{m_x m_y}$, where $m_{x/y}$ is the effective mass along $x/y$ direction, this will make the DOS around the two VHSs different, which can be seen in Fig. S1. The saddle points on the $X - M$ lines (VHS2) are close to the X point and this will generate strong ferromagnetic spin fluctuations, which promoting spin-triplet pairing. This explains the dominant spin-triplet pairing near the VHS2 filling ($|t'| > 0.3$) in our fRG calculations.

## SV. TOPOLOGICAL EDGE AND CORNER STATES

In this section we study the topological edge and corner states of the $A_1$ and $A_1 + iB_1$ superconducting phases. With the spinor $\Psi_{\mathbf{k}} = (c_{\mathbf{k}\uparrow}^\dagger, c_{\mathbf{k}\downarrow}^\dagger, c_{-\mathbf{k}\uparrow}, c_{\mathbf{k}\downarrow})$ the BdG Hamiltonian can be written as $\mathcal{H}_{\mathrm{BdG}} = \sum_{\mathbf{k}} \Psi_{\mathbf{k}}^\dagger H_{\mathrm{BdG}}(\mathbf{k})\Psi_{\mathbf{k}}$ with the Hamiltonian matrix

$$H_{\mathrm{BdG}}(\mathbf{k}) = \begin{pmatrix} h(\mathbf{k}) & \Delta(\mathbf{k}) \\ \Delta^\dagger(\mathbf{k}) & -h^\star(\mathbf{k}) \end{pmatrix}, \tag{S41}$$

where the pairing term is $\Delta(\mathbf{k}) = [\Delta_S(\mathbf{k}) + \Delta_T(\mathbf{k})\vec{d} \cdot \vec{\tau}]i\tau_2$, with $\Delta_S(\Delta_T)$ being the spin-singlet (spin-triplet) pairing component. For the helical $p$-wave ($A_1$ pairing state), the pairing matrix is

$$\Delta_T^{A_1}\vec{d} = \Delta_p(-\sin k_y, \sin k_x), \tag{S42}$$

with $\vec{d}$ being parallel to $\vec{g}$ ($\vec{d}_{\mathbf{k}} \parallel \vec{g}_{\mathbf{k}}$). The singlet pairing component is

$$\Delta_S(\mathbf{k}) = \Delta_0 + \Delta_s(\mathbf{k}) + \Delta_d(\mathbf{k}) = \Delta_0 + \Delta_s(\cos k_x + \cos k_y) + \Delta_d(\cos k_x - \cos k_y). \tag{S43}$$

The BdG Hamiltonian can be written as,

$$H_{\mathrm{BdG}}(\mathbf{k}) = \epsilon_{\mathbf{k}}\eta_z\tau_0 + g_x(\mathbf{k})\eta_0\tau_x + g_y(\mathbf{k})\eta_z\tau_y - \Delta_p[d_x(\mathbf{k})\eta_x\tau_z + d_y(\mathbf{k})\eta_y\tau_z] - \Delta_S(\mathbf{k})\eta_y\tau_y, \tag{S44}$$

where $\vec{\eta}$ and $\vec{\tau}$ are Pauli matrices in the Nambu and spin space, respectively. This Hamiltonian has time-reversal symmetry, particle-hole symmetry and chiral symmetry, satisfying $\mathcal{T}H_{BdG}(\mathbf{k})\mathcal{T}^{-1} = H_{\mathrm{BdG}}(-\mathbf{k})$ with $\mathcal{T} = i\tau_2\mathcal{K}$, $\mathcal{C}H_{\mathrm{BdG}}(\mathbf{k})\mathcal{C}^{-1} = -H_{\mathrm{BdG}}(-\mathbf{k})$ with $\mathcal{C} = \eta_1\mathcal{K}$ and $\mathcal{S}H_{\mathrm{BdG}}(\mathbf{k})\mathcal{S}^{-1} = -H_{\mathrm{BdG}}(\mathbf{k})$ with $\mathcal{S} = \mathcal{TC} = i\eta_1\tau_2$. Owing to $\mathcal{T}^2 = -1$, $\mathcal{C}^2 = +1$ and $\mathcal{S}^2 = 1$, the system belongs to the DIII class of topological superconductors [3] and is characterized by a $\mathbb{Z}_2$ invariant in two dimensions [76].

To study the edge states, we expand the above model around the $\Gamma$ point. (The expansion around the M point is briefly discussed at the end.) The effective model around the $\Gamma$ point reads,

$$H_{\mathrm{BdG}}^\Gamma(\boldsymbol{q}) = [M_0 + B(q_x^2 + q_y^2)]\eta_z\tau_0 - Aq_y\eta_0\tau_x + Aq_x\eta_z\tau_y - \Delta_p(-q_y\eta_x\tau_z + q_x\eta_y\tau_z) - \Delta_S(\boldsymbol{q})\eta_y\tau_y. \tag{S45}$$

Here, the parameters are given by

$$M_0 = -4t - 4t' - \mu, \qquad B = t + 2t', \qquad A = t\lambda,$$
$$\text{and} \quad \Delta_S = \Delta_0 + 2\Delta_s + \frac{1}{2}\Delta_s\left(q_x^2 + q_y^2\right) + \frac{1}{2}\Delta_d\left(q_x^2 - q_y^2\right). \tag{S46}$$

We first omit the singlet pairing and study the topological edge states for the helical $p$-wave pairing. Considering open boundary conditions along the $x$ direction, we let $q_x \to -i\partial_x$ in the Hamiltonian (S45) and obtain

$$H_{\mathrm{BdG}}^\Gamma(-i\partial_x, q_y) = H_0^\Gamma(-i\partial_x, q_y) + H_1^\Gamma(-i\partial_x, q_y),$$
$$H_0^\Gamma(-i\partial_x, q_y) = (M_0 - B\partial_x^2)\eta_z\tau_0 - iA\eta_z\tau_y\partial_x + \Delta_p\eta_y\tau_0\partial_x,$$
$$H_1^\Gamma(-i\partial_x, q_y) = -Aq_y\eta_0\tau_x + \Delta_p q_y\eta_x\tau_z, \tag{S47}$$

where we keep the dominant linear $q_y$ terms but neglect the insignificant $q_y^2$ terms. In the following, we solve the Hamiltonian $H_0^\Gamma(-i\partial_x, q_y)$ and treat $H_1^\Gamma(-i\partial_x, q_y)$ perturbatively. As the system belongs to DIII class, zero-energy localized edge modes will appear in the topologically nontrivial regime. The eigenvalue equation for these zero modes reads $H_0^\Gamma(-i\partial_x, q_y)\psi(x) = 0$ with the boundary condition $\psi(0) = \psi(+\infty) = 0$. With the wavefunction ansatz $\psi(x) = e^{-\alpha x}\phi$, the eigenvalue equation is further simplified to

$$\left[(M_0 - B\alpha^2)\eta_z\tau_0 + iA\alpha\eta_z\tau_y - i\Delta_p\alpha\eta_y\tau_0\right]\phi = 0. \tag{S48}$$

By multiplying $\eta_z\tau_0$ from the right hand side, the above equation can be brought into the form

$$[(M_0 - B\alpha^2) + iA\alpha\tau_y - \Delta_p\alpha\eta_x]\phi = 0. \tag{S49}$$

Therefore, the four-component wavefunction $\phi$ is an eigenstate of both $\sigma_y$ and $\eta_x$, and the corresponding eigenbasis is

$$\phi_1 = |\eta_x = 1\rangle \otimes |\tau_y = 1\rangle,$$
$$\phi_2 = |\eta_x = 1\rangle \otimes |\tau_y = -1\rangle,$$
$$\phi_3 = |\eta_x = -1\rangle \otimes |\tau_y = 1\rangle,$$
$$\phi_4 = |\eta_x = -1\rangle \otimes |\tau_y = -1\rangle. \tag{S50}$$

For the four $\phi_i$ of the eigenbasis we obtain the equation $B\alpha^2 + (-iAf_\sigma^i + \Delta_p f_\eta^i)\alpha - M_0 = 0$, with $f_{\sigma/\eta}^i = \pm$, which has the following roots

$$\alpha_{1,2} = \frac{(-\Delta_p f_\eta^i + iAf_\sigma^i) \pm \sqrt{(-\Delta_p f_\eta^i + iAf_\sigma^i)^2 + 4BM_0}}{2B}. \tag{S51}$$

The boundary condition $\psi(+\infty) = 0$ implies Re$(\alpha) > 0$. With $\alpha_1 + \alpha_2 = (-\Delta_p f_\eta^i + iAf_\sigma^i)/B$ and taking $A, B > 0$, only $\phi_{3,4}$ satisfies the above constrain with $f_\eta^i = -1$. As the pairing potential $\Delta_p$ is small, we omit $A^2/B^2$ and $A/B^2$ terms in the following calculations. With including the boundary condition $\psi(0) = 0$, the two wavefunctions can be written as

$$\psi_{3/4}(x) = \mathcal{N}_0 e^{-\beta_0 x} e^{\mp i\beta_1 x} \sin(\beta_2 x)\phi_{3/4}, \tag{S52}$$

with $\beta_0 = \Delta_p/2B$, $\beta_1 = A/2B$, $\beta_2 = \sqrt{-4BM_0 + A^2}/2B$ and $\mathcal{N}_0 = \sqrt{\frac{4\beta_0(\beta_0^2 + \beta_2^2)}{\beta_2^2}}$. In the basis of $\psi_{3,4}(x)$, the elements of $H_1^\Gamma(-i\partial_x, q_y)$ can be written as

$$\begin{aligned}
\langle\psi_3|H_1^\Gamma(-i\partial_x, q_y)|\psi_3\rangle &= \langle\psi_4|H_1^\Gamma(-i\partial_x, q_y)|\psi_4\rangle = 0, \\
\langle\psi_3|H_1^\Gamma(-i\partial_x, q_y)|\psi_4\rangle &= \mathcal{N}_1(A - i\Delta_p)q_y, \\
\langle\psi_4|H_1^\Gamma(-i\partial_x, q_y)|\psi_3\rangle &= \mathcal{N}_1(A + i\Delta_p)q_y,
\end{aligned} \tag{S53}$$

with the factor $\mathcal{N}_1 = \frac{\Delta_p}{A}(1 - \frac{A^2}{4BM_0})$. Here, we used $\tau_x|\tau_y = \pm 1\rangle = \pm i|\tau_y = \mp 1\rangle$ and $\tau_z|\tau_y = \pm 1\rangle = |\tau_y = \mp 1\rangle$. Therefore, the effective Hamiltonian of the helical edge states reads

$$\tilde{H}_1^\Gamma(q_y) = \mathcal{N}_1(Aq_y s_x - \Delta_p q_y s_y), \tag{S54}$$

with $s$ being the Pauli matrices in the $\psi_{3,4}(x)$ space. The bulk topological invariant can be directly calculated using Ref. [58]. For the used parameter setting $t = 1, t' = -0.5, \mu = -1.6$, there are four Fermi surfaces: the pockets around $\Gamma$ and $X/Y$ points are attributed to the positive helicity (+) band and the pocket around the M point is attributed to the negative helicity (-) band. As the pairing potential for the pockets around $\Gamma$ and M changes sign, the system is $\mathbb{Z}_2$ topologically nontrivial according to $N_{2D} = \Pi_s[\text{sgn}(\Delta_s)]^{m_s}$ [58] and the Dirac cone in the edge states should be located at the projection of the M point, namely $\overline{X}/\overline{Y}$ in the 1D edge Brillouin zone.

Next, we consider the spin-singlet pairing perturbation $H_2^\Gamma = \Delta_S(q)\eta_y \tau_y$ with a real gap. It can be easily shown that this perturbation cannot gap out the edge states. Then we consider the $iB_1$ pairing perturbation $H_3^\Gamma(q) = \frac{1}{2}\Delta_d(q_x^2 - q_y^2)\eta_x \tau_y$, which breaks time-reversal symmetry. With open boundary conditions in the $x$ direction, we replace $q_x$ by the momentum operator and obtain $H_3^\Gamma(q) = -\frac{1}{2}\Delta_d(\partial_x^2 + q_y^2)\eta_x \tau_y$. Projecting this into the basis of $\psi_{3/4}(x)$ gives

$$\tilde{H}_3^\Gamma(q_y) = -\frac{\mathcal{N}_2\Delta_d}{2}s_z, \tag{S55}$$

with $\mathcal{N}_2 = \frac{A^2 - 4BM_0}{2B}$. Thus, the time-reversal breaking $iB_1$ pairing introduces a mass term for the edge states. Moreover, we find that the mass term changes sign between the $x$-terminated and $y$-terminated edges, due to the $d$-wave pairing nature. Hence, the edge Hamiltonians on the $x$- and $y$-terminated edges are given by

$$\tilde{H}_x^\Gamma(q_y) = \mathcal{N}_1(Aq_y s_x - \Delta_p q_y s_y) - \frac{\mathcal{N}_2\Delta_d}{2}s_z, \tag{S56}$$

$$\tilde{H}_y^\Gamma(q_x) = \mathcal{N}_1(Aq_x s_x - \Delta_p q_x s_y) + \frac{\mathcal{N}_2\Delta_d}{2}s_z, \tag{S57}$$

respectively. When these two edges meet at a corner, the mass term must go through zero. As a consequence, a Majorana zero-energy state appears at the corner, indicating higher-order topological superconductivity.

Expanding the BdG Hamiltonian (S44) around the M point, we obtain an expression of the same form as Eq. (S45) but with the modified parameters

$$\begin{aligned}
M_0 &= 4t - 4t' - \mu, \\
B &= -t + 2t', \\
A &= -t\lambda.
\end{aligned} \tag{S58}$$

With our parameter choices for $t$, $t'$ and $\lambda$, $B < 0$ and $A < 0$ but $AB > 0$ still holds. Therefore, the wavefunctions and the corresponding edge Hamilitoians remain similar. Therefore, Majorana corner modes still appear.

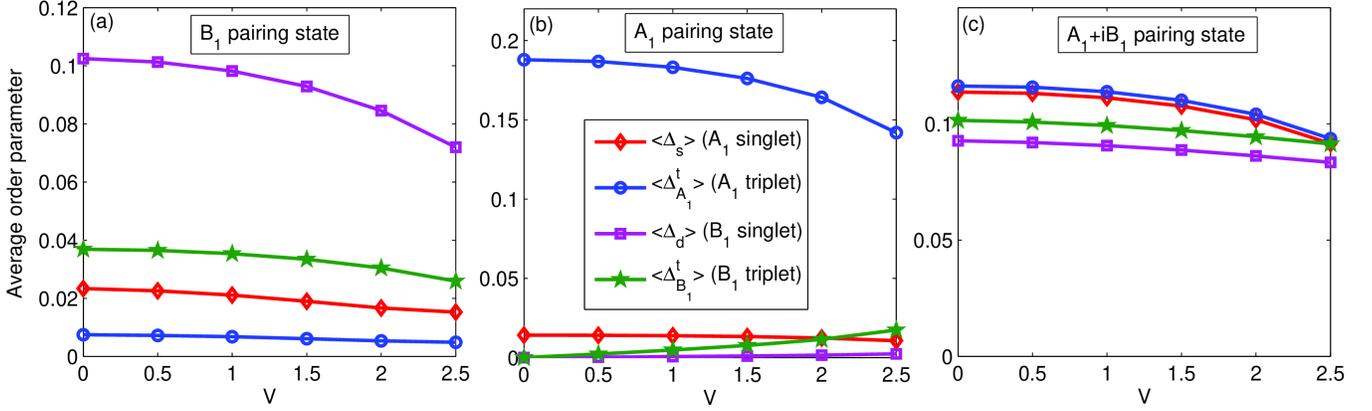

Figure S2. Dependence of average order parameter magnitudes as a function of disorder strength $V$ for three representative points of the phase diagram, (a) $\mu = -0.5$ and $t' = -0.5$ (cross I in Fig. 1 of the main text), (b) $\mu = -1.6$ and $t' = -0.5$ (cross III in Fig. 1 of the main text), and (c) $\mu = -1.2$ and $t' = -0.5$ (cross II in Fig. 1 of the main text). $V$ is expressed in units of t.

## SVI. DISORDER EFFECTS ON THE SUPERCONDUCTING STATES

Here we study the disorder effects of the three different kinds of topological superconducting states obtained in the phase diagram shown in Fig. 1 of the main text. We introduce non-magnetic chemical potential disorder by adding a term $H_V = \sum_i V_i n_i$ to the Hamiltonian with the disordered Hamiltonian given by $H_{dis} = H_{eff} + H_V$ where $H_{eff}$ is given in Eq. (S29) and $V_i$ is a site-dependent non-magnetic impurity potential drawn from a random distribution, such that $V_i \in [-V/2, V/2]$ uniformly, also commonly known as Anderson disorder. With the new method of fRG+MFT developed in this paper, we can now investigate the disorder effects of the different superconducting phases. Using the same method as in Sec. SIII, we self-consistently solve for all the order parameters with periodic boundary conditions in a $30 \times 30$ lattice. In the presence of disorder, all the order parameters vary at every site. Hence to quantify the average superconducting properties, we define average of the order parameter magnitudes as $\langle \Delta_{s/d} \rangle = 1/N \sum_i |\Delta_{s/d}|$ and $\langle \Delta_{A_1/B_1}^t \rangle = 1/N \sum_i |\Delta_{A_1/B_1}^t|$ with the site-dependent order parameters given by the definitions in Eq. (S32) and Eq. (S33). In Fig. S2, we show the disorder-dependence of average order parameter magnitudes for three representative points of the phase diagram marked by three crosses in Fig. 1 of the main text with different dominant superconducting instabilities. For $V = 0$ the dominant instabilities are clearly visible in (a) and (b): for $B_1$ pairing state it is the singlet $\Delta_d$ and for $A_1$ pairing state it is the $\Delta_{A_1}^t$. In (c), the coexistence of $B_1$ and $A_1$ pairing states is also apparent with all the order parameters being comparable in magnitude. With increasing disorder, there is a reduction in the dominant order parameters. However, the dominant orders survive with considerable magnitudes even upto $V = 2.5$ for all the three cases making them quite robust to disorder. Since the computational cost increases in the presence of disorder, the disorder effect on the whole phase diagram is left for a future work. We have also verified that the edge properties of all three topological state do not change atleast for $V \leq 2.5$. The results in this section are only presented for a single configuration of disorder. We do not expect the disorder-dependence of the order parameters to change after disorder averaging.



\end{document}